# Survivors and Zombies: The Quenching and Disruption of Satellites Around Milky Way Analogs

Debosmita Pathak[1,2], Charlotte R. Christensen[3], Alyson M. Brooks[4,5], Ferah Munshi[6], Anna C. Wright[5,7], and Courtney Carter[3]

[1] Department of Astronomy, Ohio State University, 180 West 18th Avenue, Columbus, OH 43210, USA; pathak.89@buckeyemail.osu.edu
[2] Center for Cosmology and Astroparticle Physics, Ohio State University, 191 West Woodruff Avenue, Columbus, OH 43210, USA
[3] Physics Department, Grinnell College, 1116 Eighth Avenue, Grinnell, IA 50112, USA
[4] Department of Physics and Astronomy, Rutgers University, the State University of New Jersey, 136 Frelinghuysen Road, Piscataway, NJ 08854-8019, USA
[5] Center for Computational Astrophysics, Flatiron Institute, 162 Fifth Avenue, New York, NY 10010, USA
[6] Department of Physics and Astronomy, George Mason University, 4400 University Drive, MSN: 3F3, Fairfax, VA 22030, USA
[7] Center for Astrophysical Sciences, William H. Miller III Department of Physics & Astronomy, Johns Hopkins University, 3400 North Charles Street, Baltimore, MD 21218, USA

## Abstract

It is necessary to understand the full accretion history of the Milky Way in order to contextualize the properties of observed Milky Way satellite galaxies and the stellar halo. This paper compares the dynamical properties and star formation histories of surviving and disrupted satellites around Milky Way–like galaxies using the D.C. JUSTICE LEAGUE suite of very high-resolution cosmological zoom-in simulations of Milky Way analogs and their halo environments. We analyze the full census of galaxies accreted within the past 12 Gyr, including both surviving satellites at $z = 0$, and dwarf galaxies that disrupted and merged with the host prior to $z = 0$. Our simulations successfully reproduce the trends in $M_*$–[Fe/H]–[$\alpha$/Fe] observed in surviving Milky Way satellites and disrupted stellar streams, indicating earlier star formation for disrupted progenitors. We find the likelihood and timescales for quenching and disruption are strongly correlated with the mass and time of infall. In particular, none of the galaxies accreted more than 12 Gyr ago survived, and only 20% of all accreted galaxies with $M_* > 10^8\ M_\odot$ survive. Additionally, satellites with highly radial trajectories are more likely to quench and disrupt. Disruption proceeds quickly for $\geqslant 10^6\ M_\odot$satellites accreted 10–12 Gyr ago, often on timescales similar to the $\sim$300 Myr snapshot spacing. For high-mass satellites, the disruption timescale is faster than the quenching timescale. As a result, 92% of disrupted galaxies remain star forming up until disruption. In contrast, ultrafaint dwarfs (UFDs) tend to quench prior to accretion, and 94% of UFDs accreted up to 12 Gyr ago survive at $z = 0$.

## 1. Introduction

The hierarchical assembly of the Milky Way is generally investigated through two different observational channels—analysis of the substructure within the stellar halo (e.g., A. Helmi et al. 2018; G. C. Myeong et al. 2019) and studies of surviving satellite galaxies (e.g., D. R. Weisz et al. 2015; S. P. Fillingham et al. 2019). In the former, observers identify halo substructure and piece together the assembly history of the Milky Way by leveraging the similar kinematics and chemical abundance patterns expected of stars born in the same environment (see, e.g., K. Freeman & J. Bland-Hawthorn 2002; A. Helmi 2020; A. Bonaca & A. M. Price-Whelan 2025, for recent reviews).

For example, recent evidence from combining large-scale astrometric and spectroscopic surveys such as Gaia-ESO (Gaia Collaboration et al. 2016) and APOGEE (S. R. Majewski et al. 2017), points to a major merger at $z \approx 2$ that deposited a large proportion of the stars currently in the inner halo of the Milky Way, Gaia–Sausage–Enceladus (GSE; e.g., V. Belokurov et al. 2018; M. Haywood et al. 2018; A. Helmi et al. 2018; H. Koppelman et al. 2018, and references therein). Further mapping of stellar streams has revealed a complicated history of merger events; for example, R. P. Naidu et al. (2020) attribute $\sim$95% of all halo stars between 15 and 50 kpc of the Galactic center to be of ex situ origin—a combination of accreted substructure and dynamically heated disk stars. In a complementary method, the accretion history of the Milky Way satellites can be determined by inferring orbital histories from their kinematic measurements (e.g., H. Lux et al. 2010; G. W. Angus et al. 2011; M. Rocha et al. 2012). In some cases, these two techniques can be combined by associating the phase space characteristics of satellite galaxies, globular clusters, and stellar streams to identify merging groups, as in K. Malhan et al. (2022) and F. Hammer et al. (2023).

Despite these different approaches, there is an intrinsic connection between the population of progenitor galaxies that built the stellar halo and present-day satellite galaxies: satellite galaxies represent the subset of accreted subhalos that have not yet been disrupted. As such, they are the tip of the iceberg, and by some estimates represent only $\sim$25%–30% of the total accreted satellite population with stellar mass $M_* > 10^5\ M_\odot$ (see, e.g., A. Fattahi et al. 2020; I. B. Santistevan et al. 2020). Therefore, the population of surviving satellites is a function of both the properties of the entire accreted sample of subhalos and the likelihood of disruption for any given accreted subhalo. For example, the tidal stripping of massive satellites (made more fragile by feedback-induced cores) can explain why observed satellites have lower central masses than predicted by dark-matter-only simulations (A. Zolotov et al. 2012). Similarly, the disruption of satellite galaxies has been put forth to explain the relative lack of substructure in the center of the Milky Way (S. Garrison-Kimmel et al. 2017), which decreases the likelihood of randomly placed satellites forming a plane (S. H. Ahmed et al. 2017).

There is some indication that the star-forming properties of surviving Milky Way satellites differ from the satellites of other $L_*$ galaxies in important ways. In particular, the Milky Way is a $1\sigma$ outlier with a larger quenched fraction of satellites with $10^{7.5}\,M_\odot < M_* < 10^{8.5}\,M_\odot$ compared to that of the Milky Way–mass hosts observed in the Satellites Around Galactic Analogs (SAGA; M. Geha et al. 2017, 2024; Y.-Y. Mao et al. 2021, 2024), which is a spectroscopic survey characterizing the star-forming properties of the bright and classical dwarf satellites ($10^{6-10}\,M_\odot$) of 101 relatively isolated Milky Way–like hosts. The Local Group quenched fraction is somewhat more similar to that detected by the Exploration of Local Volume Satellites (ELVES) survey, a volume-limited, photometric survey of the satellites of 30 Milky Way–like hosts in the Local Volume (S. G. Carlsten et al. 2022), which probes satellite masses of $M_*/M_\odot \gtrsim 5 \times 10^5$. Specifically, when considering all satellites with $5.5 < \log(M_*/M_\odot) < 9.5$, the quenched fractions of satellites in the ELVES survey are consistent with those of the Milky Way. However, the Local Group shows a quenched fraction on the high end of the ELVES survey for moderate-mass satellites of $7.5 \lesssim \log(M_*/M_\odot) \lesssim 8.5$ (S. G. Carlsten et al. 2022). Simulations of Milky Way analogs, in turn, tend to produce quenched fractions consistent with the Local Group (e.g., S. P. Fillingham et al. 2016; C. M. Simpson et al. 2018; H. B. Akins et al. 2021; G. D. Joshi et al. 2021). It is possible that some or all of this difference between the Local Group and other similar mass galaxies may be explained by the Local Group's accretion history. To understand how the specific accretion history will impact both the number of surviving dwarf satellites and the fraction that are quenched, a complete census of all accreted galaxies is required.

While the star formation histories (SFHs) and quenching timescales of surviving Milky Way satellites can be determined more or less directly from stellar population models (e.g., D. R. Weisz et al. 2014), the SFHs of satellites that merged with the Milky Way must be extrapolated from chemical abundances. As put forth by B. Robertson et al. (2005) and A. S. Font et al. (2006), the disrupted progenitor galaxies that primarily comprise the stellar halo were relatively massive and accreted at early times, resulting in the enhanced [$\alpha$/Fe] abundances observed for stars in the stellar halo (K. A. Venn et al. 2004). Chemical abundance measurements of the GSE stars that dominate the stellar halo are consistent with this interpretation. They indicate that GSE had one major star formation episode that ended when the galaxy accreted ~10 Gyr ago (C. Gallart et al. 2019; F. Vincenzo et al. 2019; S. Hasselquist et al. 2021), although recent observations by J. W. Johnson et al. (2023) point to a delay of ~1.5–2 Gyr between infall and quenching. Similar chemical evolution models show a 3.4 Gyr episode of star formation in the disrupted progenitor Wukong/LMS-1 (J. W. Johnson et al. 2023). In comparison to GSE, the Large and Small Magellanic Clouds (LMC and SMC, respectively), and the Sagittarius Dwarf Galaxy, all have chemistries indicative of extended and lower rates of star formation (S. Hasselquist et al. 2021; L. Fernandes et al. 2023). Recent observations by R. P. Naidu et al. (2022) of stellar streams corroborate that most disrupted satellites were formed at higher redshifts than surviving satellites. These observations further indicate that such progenitors assembled quickly in higher-density environments closer to the Milky Way, leading to their relatively iron-poor and $\alpha$-enhanced stellar populations. Measurements of the iron and $\alpha$-element abundances of M31 halo stars also indicate low metallicity and $\alpha$ enhancement (K. M. Gilbert et al. 2020; J. L. Wojno et al. 2023), implying a similar buildup of the halo from dwarf galaxies that formed their stars early.

What sets the abundance of quenched $M_* \sim 10^8\,M_\odot$ satellites of the Milky Way, and why are there differences in the SFHs between surviving and disrupted dwarf galaxies? The answers to both these questions may lie in relating the timescales of quenching and disruption for satellites. The relationship between these two timescales determines the fate of the accreted galaxy, which in turn informs the observed population of satellites and halo substructure at $z = 0$. A long history of $N$-body simulations and semianalytic models has shown that the vast majority of surviving satellites were accreted within the last 8 Gyr (e.g., A. R. Zentner & J. S. Bullock 2003; L. Gao et al. 2004; J. S. Bullock & K. V. Johnston 2005; F. C. van den Bosch et al. 2005; A. R. Zentner et al. 2005; A. P. Cooper et al. 2010; A. Zolotov et al. 2010) and that the Milky Way halo stellar mass primarily originated from $10^{10}\,M_\odot < M_{\rm vir} < 10^{11}\,M_\odot$ dwarf galaxies accreted prior to then (J. S. Bullock & K. V. Johnston 2005; A. J. Deason et al. 2016). Similar conclusions have been reached using hydrodynamics simulations (L. V. Sales et al. 2007; A. M. Brooks & A. Zolotov 2014; A. Fattahi et al. 2020) that include the effect of the baryonic disk on satellite survival and local models for star formation and chemical enrichment. These findings have been backed up by observations interpolating the orbits of Milky Way satellites (S. P. Fillingham et al. 2019). They have also been supported by observations of globular clusters (J. M. D. Kruijssen et al. 2020) and halo stars (R. P. Naidu et al. 2022) showing that the disrupted Gaia–Enceladus, Helmi streams, and Sequoia were all accreted at least ~9 Gyr ago.[8] In comparison, some other Milky Way–mass halos have stellar populations indicative of more recent, but also more massive mergers (B. Harmsen et al. 2023). For instance, M31 likely experienced a merger with an $M_* \sim 1.5 \times 10^{10}\,M_\odot$ halo 2 Gyr ago (R. D'Souza & E. F. Bell 2018; F. Hammer et al. 2018).

In addition to satellite accretion, simulations have also improved our understanding of satellite quenching by studying Milky Way analogs and their satellite systems at high resolution, using both dark-matter-only volumes that probe cosmic variance (see, e.g., S. Garrison-Kimmel et al. 2014), as well as models that include baryonic physics, realistic star formation, and feedback models (see, e.g., T. Okamoto et al. 2010; A. Zolotov et al. 2012). Ram pressure stripping appears to be the main quenching mechanism for most satellites, acting on very short timescales ($\lesssim$1 Gyr; see C. M. Simpson et al. 2018; R. C. Simons et al. 2020; H. B. Akins et al. 2021, and references therein). Ram pressure stripping is particularly efficient at lower satellite masses (J. E. Greene et al. 2023; J. Samuel et al. 2023), although reionization likely dominates

[8] Sagittarius, which was likely accreted more recently, is still in the process of disrupting (M. I. P. Dierickx & A. Loeb 2017).

**Table 1**
Properties of the Simulated Milky Way Analogs

| Resolution | Simulation | $M_{\rm vir}$ ($M_\odot$) | $M_{*,\rm sim}$ ($M_\odot$) | $M_{*,R}$ ($M_\odot$) | $R_{\rm vir}$ kpc | $N_{\rm tot}$ | $N_{\rm SF}$ | $N_{\rm Q,Host}$ | $N_{\rm Q,Indep}$ | $N_{\rm Disrupt}$ |
|---|---|---|---|---|---|---|---|---|---|---|
| Mint | `Sandra` | $2.0 \times 10^{12}$ | $1.6 \times 10^{11}$ | $9.0 \times 10^{10}$ | 267 | 77 | 6 | 10 | 24 | 37 |
| ⋯ | `Elena` | $6.9 \times 10^{11}$ | $6.8 \times 10^{10}$ | $3.7 \times 10^{10}$ | 186 | 31 | 1 | 1 | 9 | 20 |
| Near Mint | `Sandra` | $2.0 \times 10^{12}$ | $1.9 \times 10^{11}$ | $1.2 \times 10^{11}$ | 266 | 52 | 5 | 7 | 11 | 29 |
| ⋯ | `Elena` | $7.1 \times 10^{11}$ | $9.0 \times 10^{10}$ | $5.2 \times 10^{10}$ | 189 | 22 | 1 | 2 | 5 | 14 |
| ⋯ | `Ruth` | $1.1 \times 10^{12}$ | $1.0 \times 10^{11}$ | $5.3 \times 10^{10}$ | 214 | 17 | 0 | 4 | 8 | 5 |
| ⋯ | `Sonia` | $1.0 \times 10^{12}$ | $9.0 \times 10^{10}$ | $3.8 \times 10^{10}$ | 213 | 11 | 1 | 0 | 4 | 6 |

**Note.** An overview of each Milky Way analog at $z = 0$, including the resolution and name of the simulation, galaxy $z = 0$ virial mass ($M_{\rm vir}$), stellar mass calculated directly from simulation particle data ($M_{*,\rm sim}$), stellar mass calculated from the $R$-band luminosity assuming a stellar mass-to-light ratio of one ($M_{*,R}$), and the virial radius ($R_{\rm vir}$) from E. Applebaum et al. (2021) and H. B. Akins et al. (2021) for the Mint and Near Mint resolution simulations, respectively. The following satellite population statistics for all satellites accreted up to ∼12 Gyr ago are included: the total number of satellites permanently accreted by the host ($N_{\rm tot}$); the subset of satellites that survive at $z = 0$ and remain star forming ($N_{\rm SF}$); were quenched by the host ($N_{\rm Q,Host}$); quenched independent of the host ($N_{\rm Q,Indep}$); and finally the satellites that were disrupted by the host prior to $z = 0$ ($N_{\rm Disrupt}$). These populations reflect the sample selection and satellite categories introduced in Section 2.3.

for the very lowest-mass satellites (G. Efstathiou 1992; A. J. Benson et al. 2002; E. Applebaum et al. 2021). Tidal stripping increasingly contributes to gas loss at higher masses (L. Mayer et al. 2006).

Here we add to the literature on satellite quenching and disruption by comparing the star formation and resulting stellar chemical abundances for very high-resolution simulations of Milky Way galaxy satellites and progenitors using simulations that reproduce properties of surviving dwarf galaxies down to the ultrafaint regime (E. Applebaum et al. 2021; F. Munshi et al. 2021). This paper aims to directly compare the timescales of quenching and disruption across a range of satellite masses to holistically study the Milky Way's Galactic building blocks. Such a comparison enables a better contextualization of the current satellite population around the Milky Way, while improving our understanding of star formation in the progenitor galaxies prior to their disruption.

The paper proceeds as follows. In Section 2 we introduce the simulations (Section 2.1) and the halo selection criteria used throughout the paper (Section 2.3). This also includes a brief overview of how we distinguish between star-forming and quenched satellites, as well as surviving and disrupted satellites. In Section 3 we present our key result: satellite chemical abundance patterns consistent with the Local Group (Section 3.1), the impact of varying infall times and masses (Section 3.2), infall orbital trajectories (Section 3.3), and mass-loss rates (Section 3.4). In Section 4 we discuss the complex relationship between the timescales of quenching and disruption in the context of current and upcoming observations. Finally, in Section 5 we summarize our key results and conclusions.

## 2. Research Methods

### *2.1. Simulations*

We compare the surviving and disrupted satellite populations from two "Mint"-resolution and four "Near Mint"-resolution cosmological zoom-in smooth particle hydrodynamics (SPH) simulations of Milky Way–mass galaxies and their halo environments from the D.C. JUSTICE LEAGUE suite of simulations. The D.C. JUSTICE LEAGUE suite of simulations first appeared in J. M. Bellovary et al. (2018), and the properties of the host halos and satellites were further discussed in H. B. Akins et al. (2021, Near Mint resolution) and E. Applebaum et al. (2021) and F. Munshi et al. (2021; Mint resolution). All simulations were generated using the tree + SPH code CHANGA (H. Menon et al. 2015), which builds on the $N$-body gravity-tree code PKDGRAV (J. G. Stadel 2001) and SPH code GASOLINE (J. W. Wadsley et al. 2004). The simulations were computed from $z = 149$ to $z = 0$ assuming cosmological parameters from Planck Collaboration et al. (2016, $\Omega_0 = 0.3086$, $\Sigma_b = 0.04860$, $\Lambda = 0.6914$, $h = 0.677$, and $\sigma_8 = 0.77$).

The simulation initial conditions were generated using the "zoom-in" volume renormalization technique (N. Katz & S. D. M. White 1993), used to achieve high resolution in the targeted region while still including the effects of the large-scale cosmological environment. Within the resolved region, the dark-matter particle masses were 17,900 $M_\odot$ in the Mint resolution runs and 42,000 $M_\odot$ in the Near Mint resolution runs. The initial gas particle mass was 3310 $M_\odot$ (27,000 $M_\odot$), star particles are born with a mass of 994 $M_\odot$ (8000 $M_\odot$), and the gravitational spline force-softening resolution is 87 pc (174 pc) in the Mint (Near Mint) simulations. The minimum hydrodynamic smoothing length is 10% of the force-softening resolution. The resolved region was chosen to ensure that the volume within 2 $R_{\rm vir}$ of the main halo was largely uncontaminated by more massive particles (∼0.1% of the final halo mass). Since the resolved region is not spherical, however, uncontaminated halos could be found out to 2.5 $R_{\rm vir}$ of the main halo center. We limit our analysis to halos with less than 10% of their mass in the form of high-mass particles, and the vast majority of the satellites are entirely uncontaminated.

The main halos (Milky Way analogs) were selected from a $50^3$ Mpc$^3$ dark-matter-only volume to span a wide range of merger histories and spin parameters. E. Applebaum et al. (2021) and H. B. Akins et al. (2021) present the formation histories and $z = 0$ parameters of the two Mint and four Near Mint resolution hosts, respectively, which we summarize in Table 1. Note that the Mint sample consists of higher-resolution versions of the simulations included in the Near Mint sample, `Sandra` and `Elena`. For each simulation, the highest-resolution region at $z = 0$ extends out to ∼2 $R_{\rm vir}$ of the main halo, although we were able to identify a few satellites out to ∼2.5 $R_{\rm vir}$, since the highest-resolution region is not spherical.

### 2.2. Code Description

CHANGA models Kelvin–Helmholtz instabilities in shearing flows by using the geometric mean density in the SPH force expression (J. W. Wadsley et al. 2017). This method minimizes the numerical surface tension associated with density discontinuities. This method is critical to correctly modeling the shocks and instabilities in satellite halo gas particles as they pass through the host's circumgalactic medium (CGM) and is, therefore, key to modeling the gas loss rates due to ram pressure stripping (e.g., V. Quilis 2000). The simulations also allowed thermal diffusion across gas particles with a thermal diffusion coefficient of 0.03 (S. Shen et al. 2010).

The heating and cooling of gas and the associated chemical nonequilibrium abundances of H and He species are described in C. Christensen et al. (2012). A uniform time-dependent cosmic UV background adapted from F. Haardt & P. Madau (2012) was used to implement photoionization and heating rates. In this model, cosmological H II regions overlap at $z \approx 6.7$ (≈13 Gyr ago). We note, however, that this model likely ionizes and heats the intergalactic medium too early, and hence affects the thermodynamic properties of gas at $z > 6$ (J. Oñorbe et al. 2017). Dust shielding of H I and dust and self-shielding of $H_2$ was included assuming column lengths equal to the smoothing lengths of the particles. Along with H and He processes, additional cooling was provided by metal lines assuming photoionization equilibrium (S. Shen et al. 2010). Metals were diffused across particles using a subgrid turbulent mixing model with a metal diffusion constant of 0.03 (S. Shen et al. 2010). Oxygen and iron species are tracked independently using theoretical yields from S. E. Woosley et al. (2002) for Type II supernovae (SNe II) and F. K. Thielemann et al. (1986) for Type Ia supernovae, and modeling mass-loss rates from stellar winds following V. Weidemann (1987).

Star formation was implemented probabilistically based on the local $H_2$ abundance, gas density, and gas temperature, as described in C. Christensen et al. (2012). This formation occurs at and is recorded on 1 Myr time steps, enabling the analysis of star formation on timescales shorter than the snapshot spacing. Star formation was allowed only for particles with density $> 0.1$ amu cm$^{-3}$ and temperature $< 10^3$ K. However, due to the dependence on $H_2$ abundance, most stars formed at much higher densities. The transfer of energy from SNe II as thermal energy to surrounding gas particles was calculated using the "blast wave" subgrid model based on the properties of the local gas (G. Stinson et al. 2006) assuming $1.5 \times 10^{51}$ erg per supernova. This transfer of energy between supernovae and the interstellar medium (ISM) constituted the entirety of the stellar feedback in our simulations. Blast wave feedback in similar simulations has been shown to produce mass-loading factors consistent with energy-driven winds (C. R. Christensen et al. 2016).

The formation and growth of supermassive black holes (SMBHs) were implemented as described in J. Bellovary et al. (2011), and the procedures for modeling the mergers and feedback from SMBHs were as described in M. Tremmel et al. (2017). However, none of the SMBHs in the low-mass galaxies had accretion rates high enough to significantly influence the star formation rates (SFRs) in the galaxies (J. M. Bellovary et al. 2018).

Finally, as in C. Wheeler et al. (2019) and E. Applebaum et al. (2021), we impose a metallicity floor of $[Fe/H] = -4$ for all star particles to account for the lack of Population III enrichment models in the simulation. This is consistent with other simulations that alternatively impose a primordial metallicity floor of −4 to gas particles or include explicit treatment of Population III stars (e.g., M. Jeon et al. 2017; X. Ma et al. 2018; O. Agertz et al. 2020)

### 2.3. Sample Selection

We select individual galaxies from each snapshot (spaced 300 Myr apart) using AMIGA's Halo Finder (AHF; S. R. Knollmann & A. Knebe 2009), which identifies regions of dark-matter overdensities and assigns halo identifiers to gravitationally bound particles. We define the virial radius as the radius at which the enclosed density drops below 200 times the critical density at that redshift. We constructed merger trees for each D.C. JUSTICE LEAGUE simulation by using the database-generating software The Agile Numerical Galaxy Organization System (A. Pontzen & M. Tremmel 2018) to track particles across snapshots and determine global halo properties. Further analysis is done using the data-analysis software PYNBODY (A. Pontzen et al. 2013).

The main progenitor for each halo is traced back from $z = 0$ by selecting the halo in each previous snapshot that contained the majority of the simulation particles from that halo. "Satellite" galaxies of the Milky Way analog in each simulation are defined to be those halos that cross within the virial radius of its main progenitor at any time, excluding backsplash galaxies but including all galaxies accreted onto the Milky Way analog and disrupted. Excluding backsplash satellites, which necessarily spent limited time close to the host's halo, reduces variability due to environments other than the host's halo in our results. Our analysis thus includes all galaxies accreted onto each host galaxy that either disrupted before $z = 0$ or survived at $z = 0$ within the virial radius of the host (observed population). To remain well above the resolution limits of the simulations, we limit our analysis to satellites with $M_* > 100\ M_\odot$ ($M_* > 5000\ M_\odot$) and $M_{vir} > 10^5\ M_\odot$ ($M_{vir} > 10^8\ M_\odot$) for the Mint (and Near Mint) resolution simulations (E. Applebaum et al. 2021). We additionally only consider satellites that consistently host stars for at least 500 Myr prior to accretion. Doing so eliminates spurious dark-matter-only halo identifications. This check automatically makes it impossible to identify satellites during the first ∼500 Myr of the simulations. So for consistency, we do not include any halos from the first 1 Gyr of the simulations in our sample, and our results are only valid for satellites accreted over the past ∼12 Gyr. For a final consistency check, we interpolate all satellite orbits, stellar masses, and SFR of each of our satellites between snapshots to look for smooth tracks in orbital trajectories and mass evolution to confirm our satellite selection. With these selection criteria, we reject 100% of AHF-identified halos during the first gigayear (before $z \sim 5$), and thereafter reject 5%–10% of AHF-identified halos as spurious halos. For the two Mint simulations for example, out of a total 146 AHF halos, this accounts for 17 halos rejected during the first gigayear, 20 rejected for not hosting any stars (i.e., dark-matter-only halos) thereafter, and finally one halo rejected for hosting a single star particle for one snapshot. The additional checks for consistent position and velocity tracks did not flag any additional halos for removal, but helped identify two high-mass halos that were "lost" by AHF for one time step and "found" after passing through the disk of the Milky Way host, which we discuss in Section 4.2. The final satellite sample is summarized for each simulation in Table 1.

To effectively compare populations of satellites, we divide our sample into a few broad categories. As in H. B. Akins et al. (2021), we use an observationally motivated quenching threshold of $2 \times 10^{-11}$ yr$^{-1}$ in specific SFR (sSFR) measured over the last 300 Myr to distinguish between star-forming and quenched satellites.

1. Surviving satellites that remain within the virial radius of the host at $z = 0$. Surviving satellites are further classified according to their SFRs and SFHs as follows.
    1. Star-forming satellites with sSFR $\geqslant 2 \times 10^{-11}$ yr$^{-1}$ at $z = 0$.
    2. Quenched satellites that transition from star forming to sSFR $< 2 \times 10^{-11}$ yr$^{-1}$ at $r \leqslant 2 \times R_{\rm vir}$ of the host. These satellites were quenched by the host galaxy.
    3. Quenched satellites that transition from star forming to sSFR $< 2 \times 10^{-11}$ yr$^{-1}$ farther than $2 \times R_{\rm vir}$ out from the host. These satellites were quenched independently of the host galaxy before any close interactions took place.
2. Disrupted satellites that merge with the host before $z = 0$.

We note that our definition of satellites quenched by the host versus quenched independently is robust. Instead of using the distance from the host at the time of quenching, if we defined quenched independently of the host as all satellites that quench more than 1 Gyr before infall (time of first $R_{\rm vir}$ crossing), only one Near Mint satellite that infell 11 Gyr ago shifts categories from quenched independently to quenched by the host. In addition to the early infall time, this satellite infell on a highly radial orbit. Since our quenched subcategories remain virtually unchanged under both definitions, our characterization of host-induced and independent quenching is statistically robust to reasonable changes in how these categories are defined. We add the caveat, though, that galaxies identified as being quenched by the host may have been subject to preprocessing. Such galaxies could have interacted with other low-mass galaxies prior to accretion onto the main host in ways that assist and speed their eventual quenching. Such preprocessing would add to the scatter in the quenching timescales. Since quenching, when it occurs, though, generally takes place within 2 Gyr of infall, small changes to this timescale should have little effect on trends in the final status of the galaxy.

We define disruption as the last time step a halo was separately identified by AHF, before merging with the central host. Further discussion on how the definition of disruption affects our results is included in Section 4.2. Unlike surviving satellites, we do not separate disrupted satellites into star forming and quenched. We tried several ways of quantifying the SFHs of disrupted satellites, including using the SFR at infall, at disruption, and using the stellar mass assembly history to find the time at which a large fraction of the stellar mass had been assembled prior to disruption. We find that nearly all disrupted satellites were actively star forming until the point of disruption, as discussed further in Section 3.3.

### 2.4. Global Satellite Properties

We compared the satellite properties for the Near Mint runs in H. B. Akins et al. (2021) and the Mint resolution runs in E. Applebaum et al. (2021). Both sets of analysis verified that the satellite populations for all hosts are consistent with the observed $z = 0$ luminosity functions for Milky Way–mass galaxies. In particular, the most massive and satellite-rich host, `Sandra`, is most similar to M31 or the more massive M81 and Centaurus A systems. In contrast, the least massive host, `Elena`, has the fewest satellites and is similar to M94. The stellar mass–halo mass relation for all galaxies is also consistent with observations (for a detailed analysis, see F. Munshi et al. 2021; C. R. Christensen et al. 2024). H. B. Akins et al. (2021) further demonstrated that the quenched fractions and quenching timescales of the Near Mint sample were consistent with the Milky Way and Andromeda. E. Applebaum et al. (2021) directly compared the kinematics of the satellite populations to several current and comprehensive satellite galaxy catalogs. They confirmed that satellites from the Mint sample reproduce the observed size–luminosity relationship from A. W. McConnachie (2012) and R. R. Muñoz et al. (2018), and that the Near Mint versions of `Sandra` and `Elena` did so for galaxies with $M_V < -8$. Similarly, the velocity dispersions and mass-to-light ratios within the half-light radii are consistent with observations from A. W. McConnachie (2012) and J. D. Simon (2019). Finally, they find the Near Mint sample matches the stellar mass–metallicity relationship with evidence for a likely under-enrichment of iron for galaxies with $L_V \lesssim 10^4\ L_\odot$. We further analyze and discuss these metallicity measurements, now including disrupted satellites, in Section 3.1.

## 3. Results

### 3.1. Satellite Metallicities

Stellar metallicities provide one of the clearest observational indicators of the SFHs of satellites, delineating between satellites with earlier and later star formation. Hence, in this section, we first check for consistency between our simulations and observations of nearby satellite galaxies. The difference in the observed [$\alpha$/Fe] for Milky Way halo stars compared to stars in surviving satellites (e.g., K. A. Venn et al. 2004) motivates the theory that the disrupted progenitors that formed much of the stellar halo had significantly earlier star formation than present-day (surviving) satellites of similar iron abundance [Fe/H] (e.g., B. Robertson et al. 2005). The higher [O/Fe] abundance in satellites is a result of their rapid star formation—stars in these progenitors were enriched primarily by SNe II, which produce greater amounts of $\alpha$ elements relative to iron. While the majority of stars in the Milky Way's stellar halo are thought to originate from the relatively massive GSE progenitor (A. Helmi et al. 2018), the analysis of nine disrupted dwarf galaxies with $M_* \approx 10^6$–$10^9\ M_\odot$ also show a similar metallicity divide (R. P. Naidu et al. 2022), with disrupted dwarf galaxies being both Fe poor and [O/Fe] enhanced compared to surviving satellites.

Figure 1 shows where the surviving satellites and disrupted progenitors from these simulations lie in stellar mass–[Fe/H]–[$\alpha$/Fe] space. As predicted, these two different populations largely occupy separate regions, which we quantify though logistic regression. When limiting the analysis to those progenitors with $M_* > 10^5\ M_\odot$, we find that the populations may be separated using a model based on $\log(M_{*,\rm in})$, [Fe/H], [O/Fe], and combinations thereof.[9] When a threshold value of $P = 0.5$ is

[9] For example:

$$\begin{aligned} P \;=\; & -12.34 + 3.21\log(M_{*,\rm in}) - 13.03[\mathrm{Fe/H}] \\ & +146.95[\mathrm{O/Fe}] + 2.19\log(M_{*,\rm in})[\mathrm{Fe/H}] \\ & -25.53\log(M_{*,\rm in})[\mathrm{O/Fe}] + 39.56[\mathrm{Fe/H}][\mathrm{O/Fe}] \\ & -5.21\log(M_{*,\rm in})[\mathrm{Fe/H}][\mathrm{O/Fe}]. \end{aligned}$$

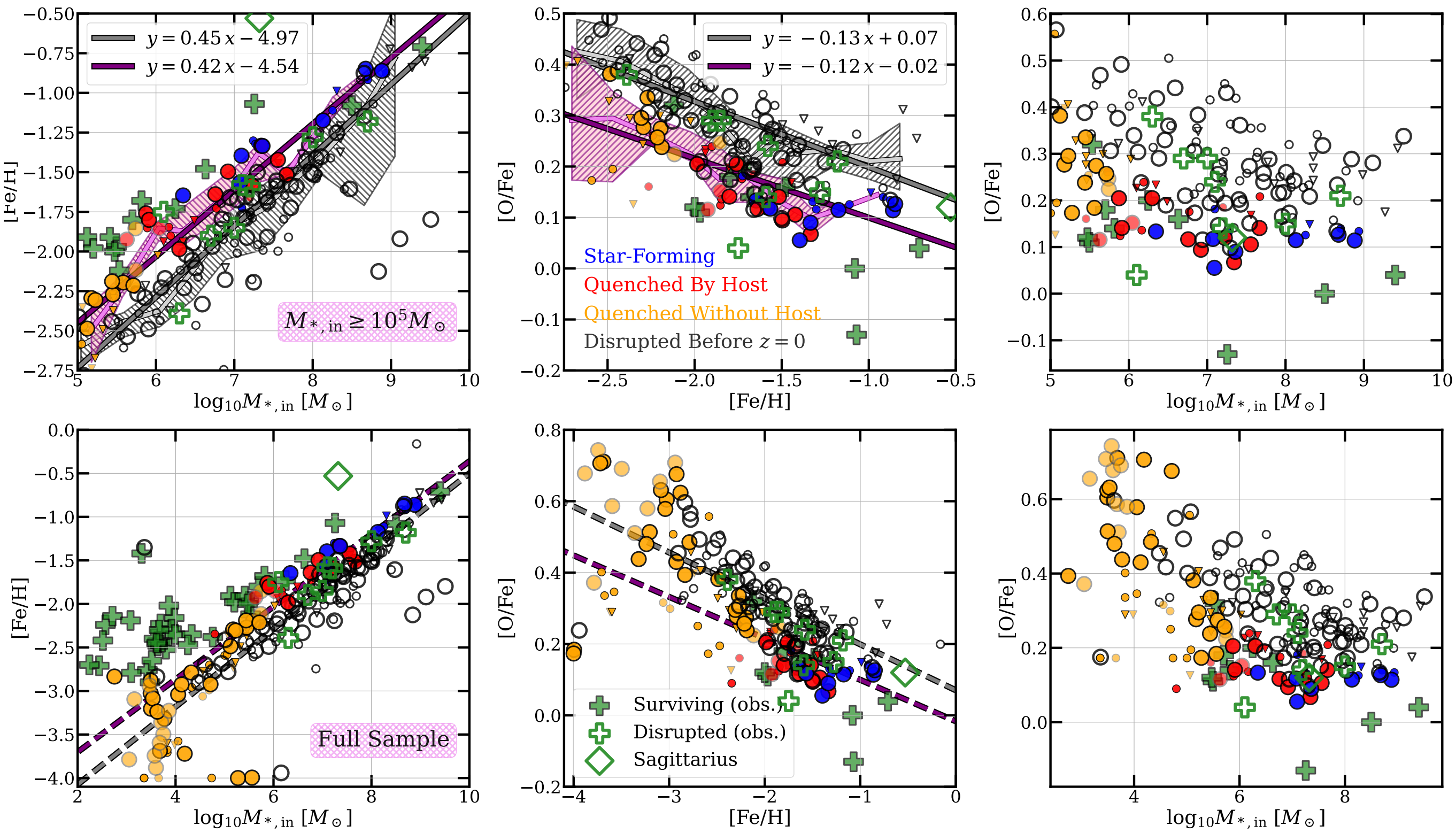

**Figure 1.** Top: the stellar mass–$Z$–$\alpha$ relation for simulated satellites in the intermediate- to high-mass range ($M_{*,\mathrm{in}} \geqslant 10^5\ M_\odot$) where satellites are colored by quenching and survival status. Disrupted satellites are shown in gray open symbols. Surviving star-forming satellites are shown in filled blue symbols, satellites quenched by the host are in red, and satellites quenched >1 Gyr prior to infall are in orange. Surviving galaxies that have fewer than 95% of their infall stars and thus likely to be in the process of tidal disruption are marked by transparent colored markers. Mint resolution satellites, Near Mint `Sandra` and `Elena`, and Near Mint `Ruth` and `Sonia` satellites are shown in large circles, small circles, and small triangles, respectively. Observed satellites are included in green, where stellar masses from surviving satellites are derived from J. D. Simon (2019) using a mass-to-light ratio of one and $M_{V,\odot} = 4.8$, and the [Fe/H] values are also from J. D. Simon (2019). Disrupted satellites (green open pluses) and Sagittarius (green open diamond) are compiled from R. P. Naidu et al. (2022). All [O/Fe] values from the observations are set to the [$\alpha$/Fe] values from R. P. Naidu et al. (2022). For all surviving and disrupted satellites $\geqslant 10^5\ M_\odot$, we provide the binned medians (light pink and gray lines, respectively), 16th–84th percentile range (pink and gray hatches), and the corresponding least squares fit to the binned medians (dark purple and gray lines). Bottom: similar to the top row, but for our full sample of satellites including observed ultrafaint dwarfs (UFDs). Our simulations appear consistent with the overall $M$–$Z$–$\alpha$ trend for satellites around the Milky Way, and successfully reproduce the relatively metal-poor but $\alpha$-enhanced population of disrupted satellites. The same best-fit lines for satellites $\geqslant 10^5\ M_\odot$ are replotted for comparison.

used to distinguish survivors from disrupted galaxies, we achieve an accuracy of 0.886 within our sample of simulated galaxies, with a sensitivity of 0.873 for identifying survivors, and a specificity of 0.895.[10] When the model trained on the simulated galaxies was applied to the sample of 19 observed galaxies with stellar masses, and [Fe/H] and [O/Fe] measurements, 17 of the galaxies were correctly classified for an accuracy of 0.895; two disrupted galaxies were mistakenly classified as survivors. The accuracy in distinguishing surviving from disrupted galaxies for both observed and simulated galaxies indicates that these populations are chemically distinct.

Disrupted satellites show higher [Fe/H] compared to their disrupted counterparts at the same stellar mass, and lower [O/Fe] compared to their disrupted counterparts in the same metallicity range. This is evident from the difference in intercept between the best-fit lines for the surviving and disrupted populations. A least squares fit to the binned data for surviving satellites results in best-fit lines of

$$
\begin{aligned}
[\mathrm{Fe/H}] &= 0.42 \times \log_{10}(M_{*,\mathrm{in}}) - 4.54 \quad (R^2 = 0.96),\\
[\mathrm{O/Fe}] &= -0.12 \times [\mathrm{Fe/H}] - 0.02, \quad (R^2 = 0.86),
\end{aligned}
$$

and for disrupted satellites:

$$
\begin{aligned}
[\mathrm{Fe/H}] &= 0.45 \times \log_{10}(M_{*,\mathrm{in}}) - 4.97 \quad (R^2 = 0.98),\\
[\mathrm{O/Fe}] &= -0.13 \times [\mathrm{Fe/H}] + 0.07. \quad (R^2 = 0.87).
\end{aligned}
$$

Surviving satellites show on average 0.3 dex higher [Fe/H] than disrupted satellites within the same stellar mass range. Similarly, surviving satellites show [O/Fe] abundances about 0.1 dex lower than their disrupted counterparts of similar [Fe/H] and stellar mass.

As summarized in the top row of Figure 1 and previously shown in E. Applebaum et al. (2021), the [Fe/H] values for our simulated surviving satellites are in good agreement with observed satellites with $M_* \gtrsim 10^5\ M_\odot$ (top row), where comparable observations are available. This trend is similar to S. E. Grimozzi et al. (2024), who compared satellites of Milky Way analogs with $M_* \gtrsim 10^6\ M_\odot$ from the ARTEMIS simulations, although our simulations show higher [Fe/H] in better agreement with observations. This agreement between our simulations and observed Milky Way satellites is an improvement over comparable simulations (e.g., A. V. Macciò et al. 2017; C. Wheeler et al. 2019; N. Panithanpaisal et al. 2021; S. E. Grimozzi et al. 2024), which tend to underpredict iron abundances (see, e.g., M. Sanati et al. 2023, and references therein), likely because of some combination of uncertainty in yields (e.g., M. S. Peeples et al. 2014; D. H. Weinberg et al. 2024), too powerful stellar feedback that ejects more [Fe/H] (e.g., O. Agertz et al. 2020; A. Fattahi et al. 2020; N. Panithanpaisal et al. 2021), lack of enrichment from Population III stars (e.g., Y. Revaz & P. Jablonka 2018), or the need for better modeling of Type Ia supernova delay times (I. Escala et al. 2018).

[10] The accuracy of this model is largely insensitive to the choice of threshold: the highest accuracy we were able to achieve was 0.905, produced by setting the threshold to 0.4184. Training the model on progenitor galaxies of all stellar masses similarly had little effect, reducing the accuracy by a couple of percent to 0.867.

Furthermore, our simulations are broadly in agreement with the observed mass–metallicity–$\alpha$ ($M$–$Z$–$\alpha$) relations between surviving and disrupted satellites, as summarized in Figure 1. Our simulations confirm that disrupted satellites show higher [O/Fe], consistent with earlier and more rapid star formation in disrupted progenitors than their surviving counterparts, as discussed further in Section 3.3. Sagittarius is a special case, indicated using a different symbol since it is still in the process of being tidally disrupted and forming stellar streams while retaining a core, and hence does not clearly fit into our definition of either surviving or disrupted satellites. Among the surviving satellites, star-forming satellites show slightly higher [Fe/H] than quenched galaxies of similar mass, consistent with S. E. Grimozzi et al. (2024), who show surviving satellites with higher gas fractions also have higher [Fe/H]. The differences in metallicity between star-forming and quenched, surviving and disrupted, satellites point to a strong dependence on accretion time and stellar mass, as seen recently in the FIRE-2 simulations as well (D. Horta et al. 2023). Despite the overall strong level of agreement, we note that our simulations lack the small population of massive surviving satellites with low $[O/Fe] \lesssim 0.1$ and high $[Fe/H] \gtrsim -1.25$, that exist around the Milky Way: Fornax and the Magellanic Clouds. This lack could be an indication that the simulations are not producing enough satellites with very late SFHs.

With our higher-resolution simulations, we extend this metallicity comparison down to $M_{*,\mathrm{in}} \geqslant 5 \times 10^2\ M_\odot$, which includes UFDs ($M_{*,\mathrm{in}} < 10^5\ M_\odot$), in the bottom row of Figure 1. Since observations may classify a somewhat different sample of satellites as disrupted than AHF, especially at the low-mass end where satellites may be in the process of disruption, Figure 1 also includes lower-opacity colored points to indicate galaxies that retain fewer than 95% of their infall star particles at $z = 0$. Doing so allows us to distinguish satellites that show signs of disruption in the form of tidal streams, even though they retain a gravitationally bound core. Not only are these low-mass galaxies less likely to be fully disrupted (see the detailed discussion of the role of satellite mass in determining tidal disruption in Section 3.2), but their quenching is likely due to the cosmic UV background radiation, as opposed to the host galaxy. As a result, all low-mass galaxies, including those that show some evidence of tidal disruption, have similar abundance patterns indicative of early SFHs.

An examination of the metallicities of UFD satellites with $M_* \lesssim 10^5\ M_\odot$ reveals a discrepancy between the simulations and observed dwarf galaxies. Photometric metallicity estimates of the average metallicity for UFDs show a "floor" in the mass–metallicity relation at $[Fe/H] \approx -2.5$ dex (e.g., E. N. Kirby et al. 2013; S. W. Fu et al. 2023). Similar to other cosmological simulations of comparable resolution (e.g., A. V. Macciò et al. 2017; C. Wheeler et al. 2019; N. Panithanpaisal et al. 2021), low-mass galaxies in our simulations have lower values of [Fe/H] than observed, resulting in a steeper slope at the low-mass end of the mass–metallicity relationship (see, e.g., M. Sanati et al. 2023; B. Azartash-Namin et al. 2024, for a comparison). Specifically, simulated galaxies from this sample with $\log_{10}(M_{*,\mathrm{in}}/M_\odot) < 5$ have $-4.0 < [Fe/H] < -2.7$. While these metallicities are low compared to the observed metallicity floor, the presence of UFD galaxies with metallicities reaching up to $[Fe/H] \sim -2.7$ is an improvement over previous work (e.g., A. V. Macciò et al. 2017; C. Wheeler et al. 2019; N. Panithanpaisal et al. 2021). Additionally, E. Applebaum et al. (2021) showed when comparing the total amount of metals, not just Fe, the overall agreement between our simulations and observed satellites improves significantly, such that our simulated satellites are on average 0.1–0.2 dex higher than the observed galaxies and only one of the low-mass observed galaxies lies outside the $1\sigma$ scatter. Therefore, it is clear that our current feedback prescription allows simulated UFDs to successfully retain metals in their ISM, although it may underproduce iron. These differences argue for further tuning the models of the UV cosmic background, Type Ia supernova timing, and strength of feedback, as investigated by O. Agertz et al. (2020). However, these changes would effectively add a systematic offset to the lowest-mass galaxies and any higher-mass ones whose star formation was also halted around the time of reionization, which accounts for <5% of all progenitors, surviving or disrupted, with $M_{*,\mathrm{in}} \geqslant 10^5\ M_\odot$, making the overall mass–metallicity trend and differences between satellite populations fairly reliable. In the following sections, we thus contextualize differences in metallicity across satellite populations by quantifying the following: infall and quenching timescales (Section 3.2), orbital trajectories and mass accretion history (Section 3.3), and mass-loss history of the full census of accreted satellites (Section 3.4).

### *3.2. Satellite Mass, and Timescales of Infall and Quenching*

In this section, we directly compare the infall stellar masses, infall times, and quenching timescales of the full census of surviving and disrupted progenitors. Figure 2 summarizes the range of infall stellar masses $M_{*,\mathrm{in}}$, quenched fraction, and surviving versus disrupted fraction of our sample of satellites for the two Mint resolution simulations. This distribution is also representative of the four Near Mint simulations for satellites with $M_{*,\mathrm{in}} \gtrsim 10^5\ M_\odot$. Comparing the cumulative distribution function (CDF; left panel of Figure 2) and histograms (right panel) of $M_{*,\mathrm{in}}$ of surviving satellites (yellow, red, and blue) and disrupted progenitors (hatched gray region), it is clear that the stellar mass at infall is a primary divider between satellite survival and disruption. The vast majority of disrupted satellites infell with high stellar masses, $M_{*,\mathrm{in}} \geqslant 10^6\ M_\odot$, consistent with the fact that massive satellites experience more rapid orbital decay because of dynamical friction (S. Chandrasekhar 1943). Satellites that survive at $z = 0$, on the other hand, show a wider range of infall masses, skewed toward lower masses.

Among the overall surviving sample, stellar mass at infall is a good indicator for distinguishing star-forming and quenched satellites. Only the most massive surviving satellites with $M_{*,\rm in} \geqslant 10^7\ M_\odot$ remain star forming; intermediate-mass satellites with $M_{*,\rm in} \sim 10^6$–$10^7\ M_\odot$ are generally quenched due to the host's influence, and low-mass satellites with $M_{*,\rm in} \lesssim 10^5\ M_\odot$ quench independently of the host, likely because of the cosmic UV radiation (H. B. Akins et al. 2021; E. Applebaum et al. 2021). In fact, about 60% of the total surviving satellite population were quenched at distances $> 2 \times R_{\rm vir}$ and at >1 Gyr prior to infall.

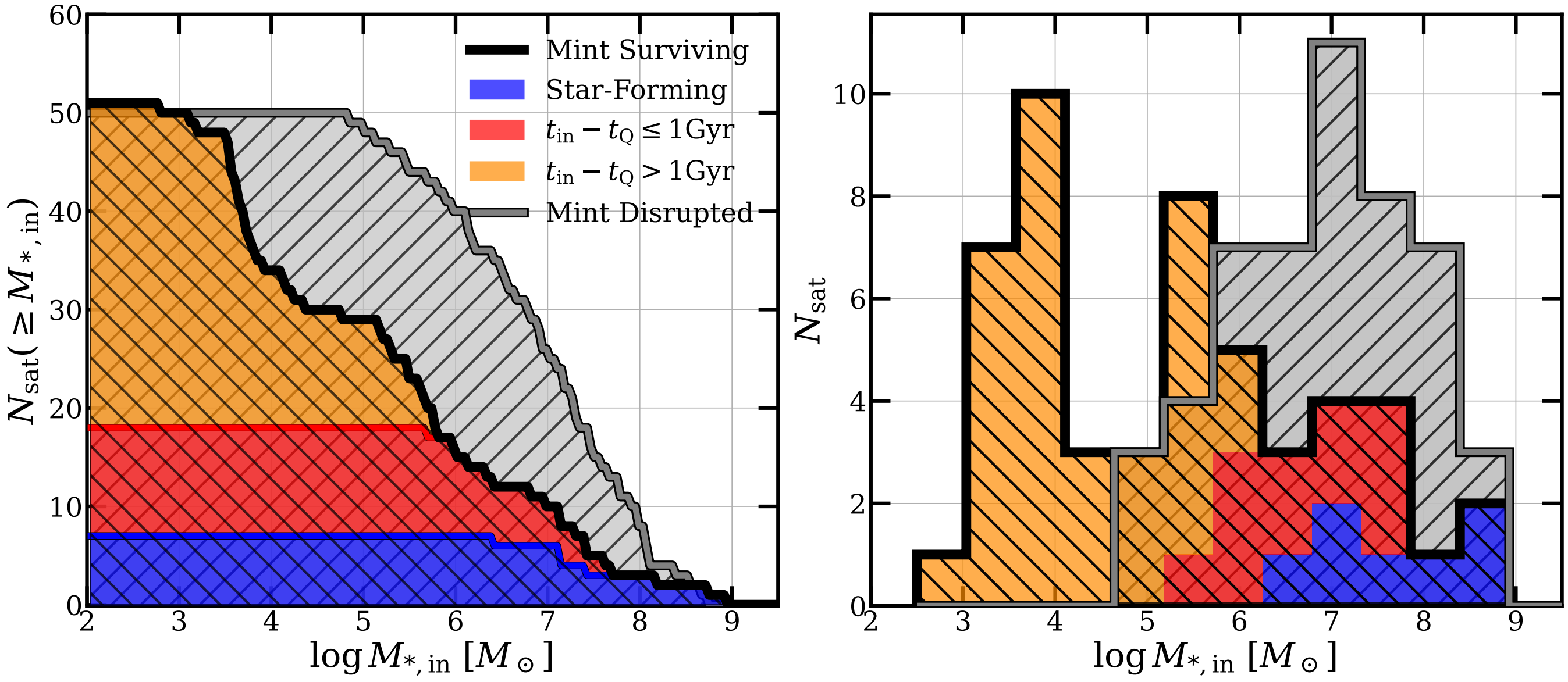

**Figure 2.** Left: CDF of infall stellar masses $M_{*,\rm in}$ of the Mint resolution sample. Using a similar coloring scheme as in Figure 1, the CDF of disrupted satellites is indicated in gray, while the contribution to surviving satellites from star-forming satellites, satellites quenched by the host, and satellites quenched >1 Gyr prior to infall are indicated in blue, red, and orange, respectively. The stacked CDFs of these three categories (indicated by the thick black line) make up the total surviving satellite population. Right: stacked histograms showing the distribution of $M_{*,\rm in}$ between surviving and disrupted satellites. We separate the surviving satellites by quenching status, as in the left panel.

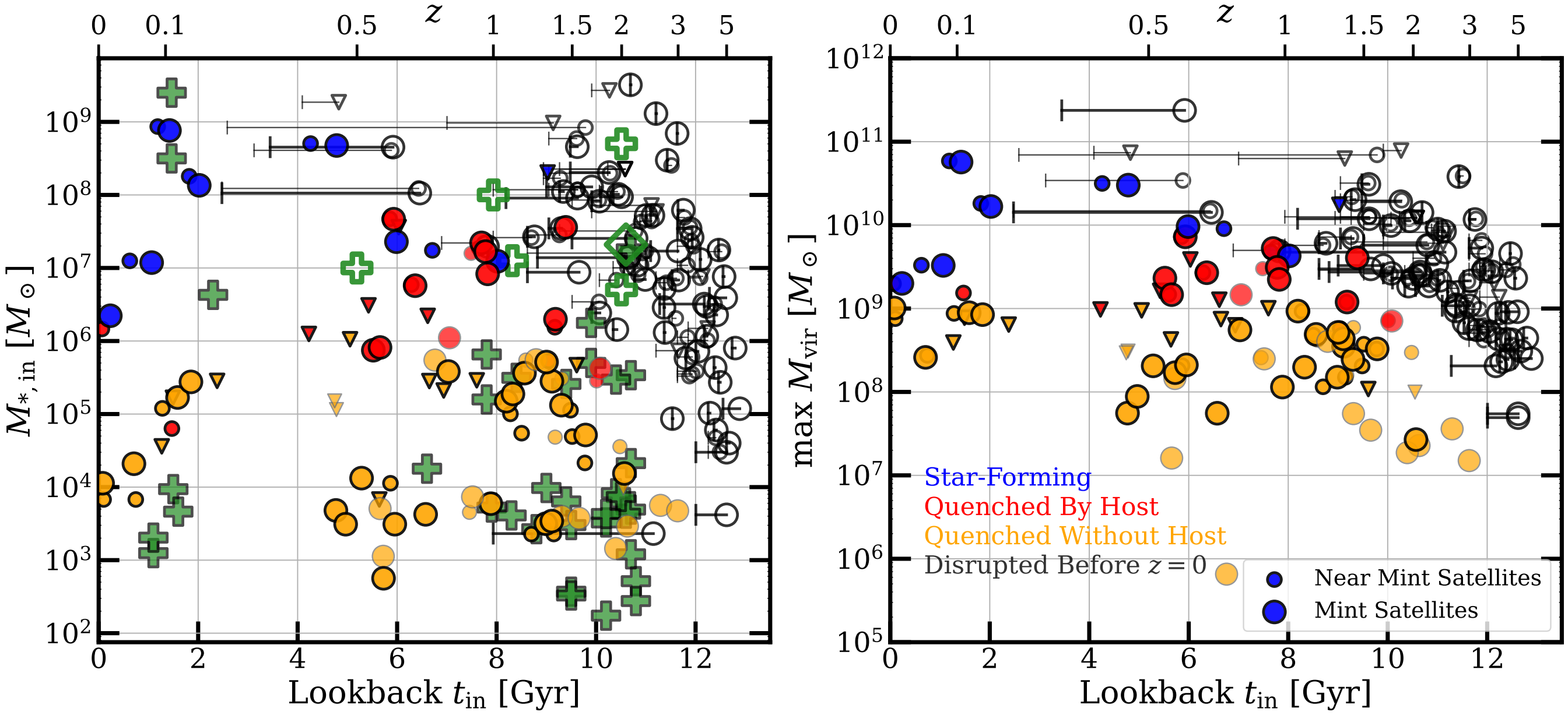

**Figure 3.** Left: stellar masses at infall ($M_{*,\rm in}$) and the lookback time to infall for our full sample, using the same color scheme as Figure 1. Horizontal bars extending out to the time of disruption are shown for each disrupted satellite (open symbols). As in Figure 1, observations are indicated by pluses, Mint resolution simulated satellites are indicated by large circles, and satellites from the Near Mint (lower-resolution) runs are indicated by smaller symbols (`Sandra` and `Elena` in circles, `Ruth` and `Sonia` in triangles). Observed surviving satellites, shown as filled green pluses, have stellar masses derived from J. D. Simon (2019) as in Figure 1, and infall times from S. P. Fillingham et al. (2019). Data for disrupted galaxies (shown as empty green pluses) are from R. P. Naidu et al. (2022). Surviving satellites that retain <95% of the stars they accreted with at $z = 0$ are indicated in lower-opacity symbols. Right: similar to the left panel, now showing the maximum virial mass of the satellites instead of the stellar mass at infall.

To break the mass degeneracy between surviving star-forming satellites and disrupted progenitors, we include the times of first infall (virial crossing) for each satellite. Figure 3 summarizes the diversity in infall times and masses captured in our full sample. The timescale of disruption within the host halo for each disrupted satellite is also indicated with gray horizontal lines connecting the infall and disruption times, $t_{\rm in}$ and $t_{\rm D}$, respectively. Observations of surviving (filled green pluses; S. P. Fillingham et al. 2019; J. D. Simon 2019) and disrupted satellites (empty pluses; R. P. Naidu et al. 2022) are included for comparison.[11] The stellar and virial masses at infall, in combination with the time of infall, are excellent indicators of the quenched and survival status of satellites. Satellites with early infall times are more susceptible to disruption by $z = 0$, since they spend longer within the host halo and were accreted at a time when the host was undergoing rapid growth and halo assembly. Except for a few low-mass satellites, very few satellites accreted more than 10 Gyr ago survive, setting a rough threshold for the earliest time of infall for surviving satellites to a lookback time of 10 Gyr. Even for lookback times of infall up to 8 Gyr ago, survival is rare for galaxies with $M_{*,\rm infall} > 10^6\ M_\odot$. These lookback times corresponding to survival or destruction are comparable to those from dark-matter-only simulations (e.g., J. S. Bullock & K. V. Johnston 2005; R. D'Souza & E. F. Bell 2021) that found median lookback accretion times of $\sim$9 Gyr for destroyed systems and $\sim$5 Gyr for surviving systems with halo masses of $>5 \times 10^9\ M_\odot$.

The oldest surviving low-mass satellites are UFDs ($M_{*,\rm in} < 10^5\ M_\odot$) accreted up to $\sim$11.5 Gyr ago. In fact, only two Mint UFDs accreted within the last 12 Gyr disrupt, and $\gtrsim$94% of UFDs accreted within the last 12 Gyr survive at $z = 0$. As discussed in Section 4.2, this fraction can change depending on how we classify satellites in the process of tidal disruption. For example, if we consider UFDs that retain fewer than 95% of their stars at infall (low-opacity points in Figure 3) as disrupted, this fraction falls to 60% of UFDs surviving. However, this translates to a relatively small fraction in overall stellar mass loss for UFDs in general, as discussed in Section 3.4. Most of these low-opacity satellites could also be characterized as on their way to being "orphan galaxies" (Q. Guo et al. 2011), since they have lost >60% of their dark matter since infall (see Section 3.4). The median satellite-to-host mass ratio for UFDs that are not disrupted at infall is $2.47 \times 10^{-7}$, and the 16th–84th percentile spread in mass ratios is $7.30 \times 10^{-8}$–$2.55 \times 10^{-6}$. On the other hand, the median satellite-to-host mass ratio for the handful of UFDs that disrupt soon after infall at $z \sim 2$ is $1.03 \times 10^{-4}$.

In summary, the vast majority of ultralow-mass satellites with $M_{*,\rm in} < 10^5\ M_\odot$ survive to $z = 0$ and quench before infall[12] (see also Figure 4). Most low- to intermediate-mass satellites with $10^5\ M_\odot < M_{*,\rm in} < 10^8\ M_\odot$ also survive at $z = 0$, and are quenched depending on their infall mass and time spent in the host halo. Finally, only the most massive satellites accreted less than 8 Gyr ago are prone to disruption before $z = 0$. Among the highest-mass satellites ($M_{*,\rm in} \geqslant 10^8\ M_\odot$), the amount of time spent in the host halo determines whether a satellite disrupts or whether it survives and remains star forming. Therefore, a combination of the mass and time of infall can largely distinguish surviving satellites from disrupted progenitors, as well as star-forming from quenched satellites.

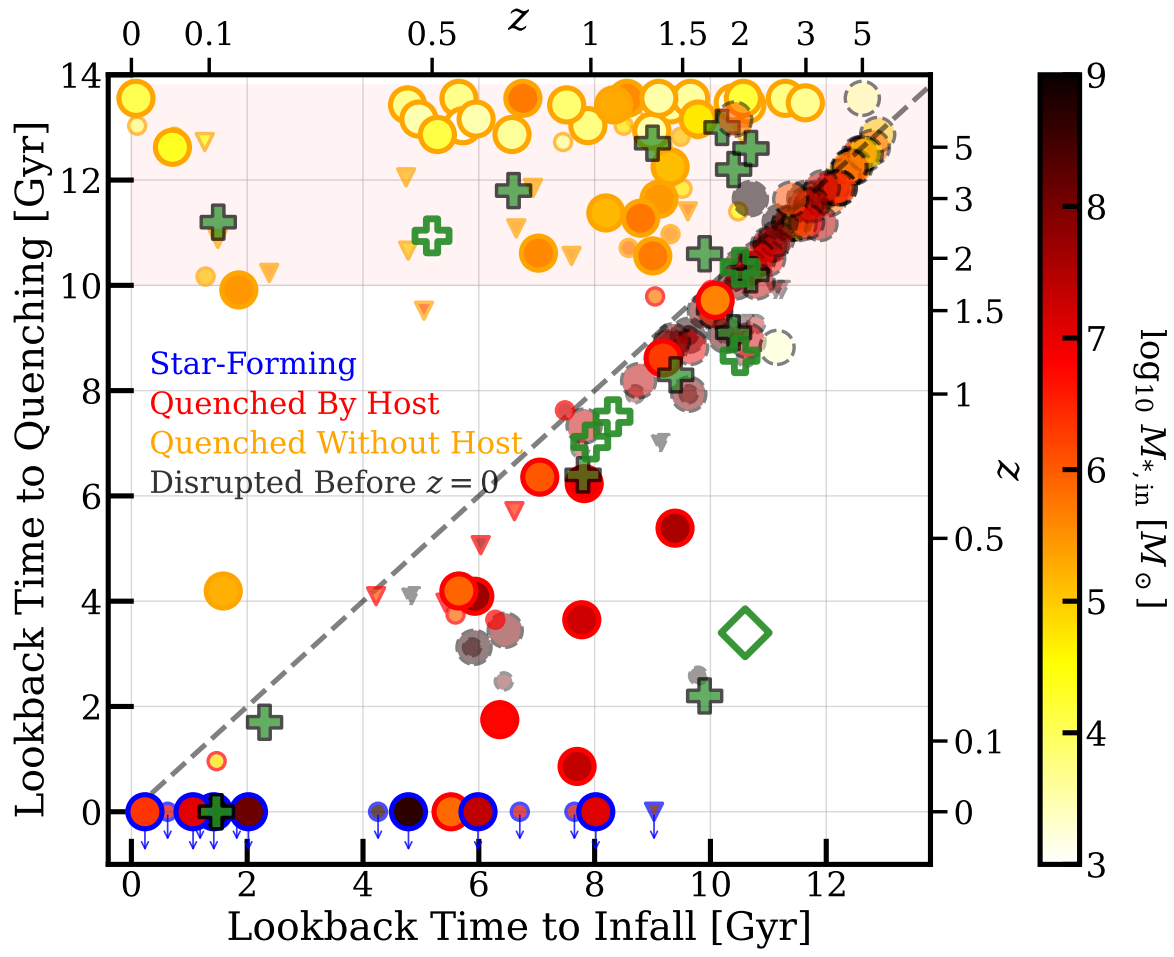

**Figure 4.** Lookback time to infall and quenching for all simulated satellites, colored by their stellar mass at infall. The top shaded region highlights the rough time of reionization. The outlines of the simulated points follow the convention from Figure 1, with disrupted satellites in fainter dashed gray outlines, and surviving star-forming, quenched by host, and quenched >1 Gyr prior to infall satellites are indicated in solid blue, red, and orange outlines, respectively. Both Mint (large circles) and Near Mint (smaller circles and triangles) are included for comparison with the observed patterns in infall and quenching time for Milky Way satellites. Observed satellites, shown as filled green pluses, are from S. P. Fillingham et al. (2019) and disrupted galaxies are from R. P. Naidu et al. (2022). Star-forming satellites are indicated at the bottom of the figure with a lookback $t_{\rm Q} = 0$ and blue arrows as a lower bound for $t_{\rm Q}$. The gray dashed line indicates the 1:1 line.

Figure 4 compares the times of infall and quenching for our sample. Observational data from S. P. Fillingham et al. (2019) and R. P. Naidu et al. (2022) are shown for comparison. The typical uncertainties in the observed infall and quenching times for surviving satellites from S. P. Fillingham et al. (2019) are large, about ±2 Gyr and ±5–10 Gyr, respectively. There is similar uncertainty in the infall and quenching time estimates of the disrupted progenitor candidates from R. P. Naidu et al. (2022), who use a combination of simulations (X. Ma et al. 2016) and inferred SFHs. Given these uncertainties, our simulations are an excellent match to the observations of both surviving and disrupted progenitors around the Milky Way. The median and 16th–84th percentile scatter in infall stellar mass, virial mass, infall time, and quenching time for all four satellite populations for the Mint simulations are summarized in Table 2.

Figure 4 also confirms that most satellites with $M_* < 10^5\ M_\odot$ (UFDs) form their stars rapidly within the first $\sim$2 Gyr of cosmic star formation and are subsequently quenched, likely by reionization (light red region). This is consistent with the relatively high [$\alpha$/Fe] and low [Fe/H] values seen in UFDs (Section 3.1). These ultralow-mass systems show a range of infall times. Most low- to intermediate-mass satellites ($10^5\ M_\odot < M_{*,\rm in} < 10^8\ M_\odot$) accreted between 5 and 10 Gyr ago are quenched within $\lesssim$3 Gyr of first entering the host's virial radius. Finally, the most massive surviving satellites ($M_* \geqslant 10^8\ M_\odot$) were accreted within the last $\sim$6 Gyr and continue to actively form stars up until $z = 0$. The typical quenching timescale for satellites quenched by the host (intermediate-mass satellites) is 1–3 Gyr after infall.

One remarkable result is that, to the best of our ability to resolve, the vast majority of disrupted galaxies with

[11] The observational estimates of $t_{\rm in}$ and $t_{\rm Q}$ show large uncertainties of order 1–10 Gyr. We hence only include the best estimate of these timescales from the literature for first-order comparison, but highlight the need for better constraints to do reliable comparisons beyond average population statistics.

[12] Many of these lowest-mass galaxies show some signs of tidal disruption, which we identify by marking those galaxies that retain fewer than 95% of the star particles they entered the host halo with at $z = 0$. Such systems may be most like the stellar streams with surviving progenitors identified in N. Panithanpaisal et al. (2021) and N. Shipp et al. (2023), and they show a similar pattern of infall times, peaking after most of the fully disrupted galaxies were accreted.

**Table 2**
Summary of Satellite Infall Properties

| Category | $\log M_{*,\rm in}$ [a] ($\log M_\odot$) | $\log M_{\rm vir,in}$ [a] ($\log M_\odot$) | Lookback $t_{\rm in}$ [a] (Gyr) | Lookback $t_{\rm Q}$ [b] (Gyr) | $\Delta t_{\rm Q}$ [b] (Gyr) | $\frac{v_{\rm rad,in}}{v_{\rm tot,in}}$ [c] |
|---|---|---|---|---|---|---|
| Surviving, star forming | $7.36_{7.04}^{8.68}$ | $9.97_{9.48}^{10.49}$ | $2.02_{1.03}^{6.07}$ | ⋯ | ⋯ | $0.48_{0.43}^{0.83}$ |
| Surviving, quenched by host | $6.76_{5.90}^{7.43}$ | $9.36_{9.10}^{9.63}$ | $7.70_{5.82}^{9.27}$ | $4.19_{1.39}^{7.26}$ | $1.84_{0.64}^{4.98}$ | $0.81_{0.63}^{0.92}$ |
| Surviving, quenched independent of host | $3.87_{3.49}^{5.45}$ | $7.89_{7.23}^{8.61}$ | $8.19_{5.00}^{9.77}$ | $13.16_{11.41}^{13.55}$ | $-4.44_{-7.86}^{-2.63}$ | $0.79_{0.53}^{0.95}$ |
| Disrupted | $7.20_{6.09}^{8.08}$ | $9.45_{8.73}^{10.11}$ | $11.10_{9.64}^{12.07}$ | ⋯ | ⋯ | $0.88_{0.79}^{0.98}$ |

**Notes.** Summary of properties for surviving satellites that remain star forming, were quenched by host, quench independently of the host, and disrupted satellites. The 50th percentile for each quantity is shown, with the 16th and 84th percentiles denoted as subscripts and superscripts, respectively.

[a] Satellite stellar mass and virial mass in units of $\log M_\odot$at the time of infall. $t_{\rm in}$ is the lookback time to infall.

[b] The lookback time to quenching $t_{\rm Q}$, and the time delay between infall and quenching, $\Delta t_{\rm Q}$. Positive $\Delta t_{\rm Q}$ values indicate quenching after infall, while negative $\Delta t_{\rm Q}$ values correspond to quenching prior to infall.

[c] The fractional radial component of total satellite velocity at infall, where higher values correspond to more radial orbits, and lower values indicate more circular orbits.

$M_{*,\rm in} > 10^5\ M_\odot$ were star forming up until the point of disruption. In other words, almost every disrupted galaxy above $\sim 10^5\ M_\odot$ was forming stars in the last snapshot it was identified (i.e., the last snapshot prior to disruption). These simulations show star formation to within 300 Myr of disruption, which is comparable to our snapshot spacing, as discussed in Sections 4.2 and 4.3. Since only four Mint satellites and one Near Mint satellite with $M_{*,\rm in} \lesssim 10^5\ M_\odot$, all of which were early accretion events, likely quenched by reionization prior to disruption, we do not divide the disrupted galaxies by their quenched status at the time of disruption. These low-mass quenched satellites comprise only 8% of the total disrupted satellite population in the Mint runs. As will be discussed further in Section 3.3, this lack of quenching among disrupted galaxies is likely because disruption typically happens rapidly in our simulations. For most low- and intermediate-mass satellites, the typical disruption timescale is <1 Gyr, or so short that we cannot resolve it at our ∼250–300 Myr snapshot resolution. For more massive satellites, the typical disruption timescale is 1–2 Gyr. In the circumstances where disruption does proceed more slowly, such as in late accreting, massive star-forming galaxies, the quenching timescales are also longer, in fact, much longer than the disruption timescale. The two timescales of satellite quenching and disruption thus compete to produce the observed population of surviving star-forming and quenched satellites at $z = 0$.

### 3.3. Orbital Dynamics

While the mass and time of infall of a satellite largely determines its disruption and quenching timescales, orbital trajectories play a secondary role in determining the fate of satellites, as seen from the larger scatter in velocities compared to masses for any given category of satellites in Figures 3 and 6. While the mass of the satellite (marker size in Figure 3) and infall time are still the primary indicators of satellite survival and quenching, when we compare between satellites of similar stellar mass that infell at similar lookback times (independently, not as group accretion events), orbital trajectories are often the deciding factor that can explain the cases of mass and infall time degeneracy where, for example, one satellite remains star forming while the other quenches (such as the two $\sim 5 \times 10^7\ M_\odot$ Mint satellites at lookback $t_{\rm in} \approx 8$ Gyr), or similarly, when one satellite survives while the other disrupts (such as the two $\sim 5 \times 10^8\ M_\odot$ Mint satellites that accreted ≈5–6 Gyr ago). This is especially relevant for satellites at the intermediate- to high-mass end, where the mass and time of infall alone are unable to explain why some satellites quench or disrupt more rapidly than others. Previous work has shown that the orbits of satellites influence their likelihood of disruption (e.g., K. V. Johnston et al. 2008; N. C. Amorisco 2017; N. Shipp et al. 2023) and of quenching (e.g., R. C. Simons et al. 2020; A. Di Cintio et al. 2021; J. Samuel et al. 2023). Here, we examine how both disruption and quenching proceed together and their sensitivity to orbital trajectory at infall.

The top row of Figure 5 shows the three-dimensional orbital trajectories of Mint resolution satellites, tracked since 1 Gyr prior to infall, grouped according to the stellar mass of each satellite at infall. Very low- ($M_{*,\rm in} < 10^4\ M_\odot$) and low-mass satellites ($10^4\ M_\odot \leqslant M_{*,\rm in} < 10^6\ M_\odot$) show a wide range of orbital trajectories after infall, most complete multiple pericentric passages, and the vast majority of these satellites survive at $z = 0$. The only two ultralow-mass satellites and one low-mass satellite that disrupt were accreted >11 Gyr ago and infell on highly radial orbits. In contrast, the highest-mass satellites ($M_* \geqslant 10^8\ M_\odot$), which are most prone to rapid disruption, survive and remain star forming if they infell on highly circular orbits and remained at large orbital radii, and disrupt rapidly within 1–2 Gyr if they infell on radial trajectories.

While the dark-matter content of subhalos can fluctuate, there is usually little reduction in stellar mass after infall. For satellites that continue to form stars until disruption, a balance between forming new stars and losing some can stabilize $M_*$ after infall, highlighting the importance of tracking baryonic content when evaluating disruption in satellites. Some high-mass satellites, in fact, show continued stellar mass growth after infall. Especially for low- and intermediate-mass satellites, most of the stellar mass at infall is retained despite satellites infalling with a range of orbital parameters and completing several pericentric passages. Most satellites are thus more resistant to stellar mass loss, even during significant loss of more loosely bound dark matter.

Comparing the circularity of orbits at infall shows differences between surviving and disrupted satellites, especially for massive satellites. Figure 6 examines the role of orbital parameters at infall for determining disruption timescales for both the Mint and Near Mint resolution samples, by comparing what fraction of the infall velocity is in the radial direction, $|v_{\rm rad,in}/v_{\rm tot,in}|$. From Figure 6, it is clear that most disrupted satellites were accreted on highly radial orbits and disrupted rapidly. On the other hand, among the massive star-forming satellites prone to disruption, only those on more circular orbits with $|v_{\rm rad,in}/v_{\rm tot,in}| \lesssim 0.6$ survive. All other massive satellites with $|v_{\rm rad,in}/v_{\rm tot,in}| \gtrsim 0.7$ disrupt within 0.5–4 Gyr of infall. Similar theoretical work has shown that highly eccentric orbits are more likely to result in stellar streams (N. Shipp et al. 2023). Therefore, in addition to the satellite mass and time of infall, satellite orbits at infall play a secondary role in influencing the likelihood of disruption. Indeed, the survival fraction of both Mint and Near Mint satellites decreases steeply as a function of $|v_{\rm rad,in}/v_{\rm tot,in}|$, as shown in Figure 6. The fraction of satellites that survive at $z = 0$ drops from ∼75% at $|v_{\rm rad,in}/v_{\rm tot,in}| \approx 0.2$ (primarily circular orbits) to <10% at $|v_{\rm rad,in}/v_{\rm tot,in}| \approx 1$ (radial trajectories). Hence, only ∼25% of satellites on circular orbits disrupt, while ∼90% of satellites on primarily radial trajectories disrupt.

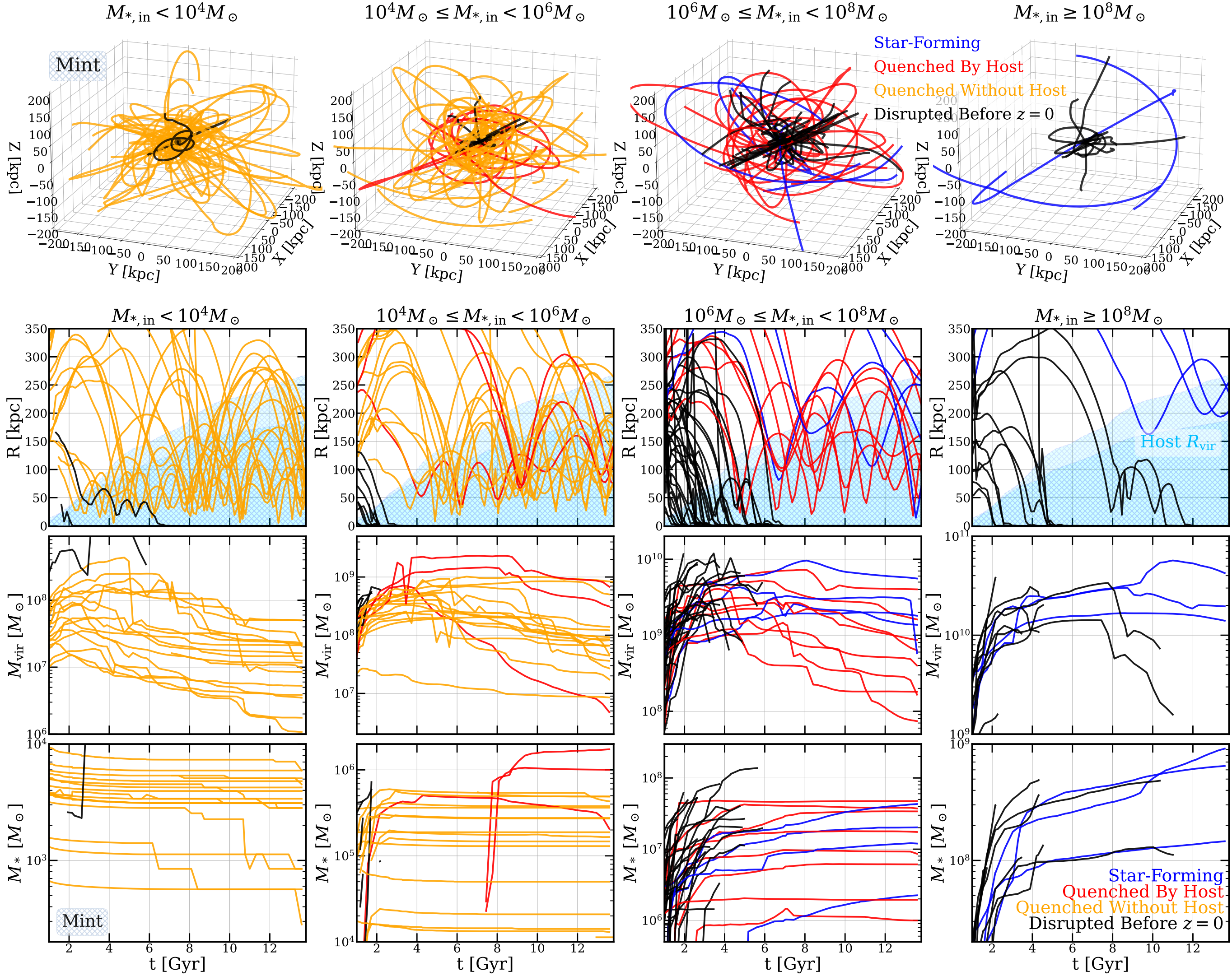

**Figure 5.** Top row: tracking satellite orbital trajectories starting 1 Gyr prior to infall for all satellites from the Mint simulations binned by stellar mass at infall. Orbits are centered on the center of mass of the host (gray dot). Black lines show the orbits of disrupted galaxies as they sink to the center and merge with the host, and solid color lines show surviving galaxies, with line colors corresponding to the convention in Figure 1. Position data for satellites are interpolated across snapshots until disruption (merger with host) for disrupted satellites and $z = 0$ for surviving satellites. From left to right, each figure corresponds to a different infall stellar mass bin: ultralow ($M_* < 10^4\ M_\odot$), low ($10^4\ M_\odot \leqslant M_* < 10^6\ M_\odot$), intermediate ($10^6\ M_\odot \leqslant M_* < 10^8\ M_\odot$), and high ($M_* \geqslant 10^8\ M_\odot$). Second row: tracking the interpolated radial distance from the central host as a function of time for all Mint resolution satellites up until disruption or $z = 0$. The evolution of the virial radii of the two Mint hosts are indicated in cyan. Third row: the virial mass of each Mint satellite up until disruption or $z = 0$. Bottom row: the stellar mass of each Mint satellite as a function of time.

Orbits play a role in quenching timescales in satellites, as satellites with radial orbits may experience more ram pressure because of close pericentric passages (e.g., R. C. Simons et al. 2020; A. Di Cintio et al. 2021; J. Samuel et al. 2023), although R. C. Simons et al. (2020) also note that CGM substructure leads to large amounts of scatter in this relationship. Among the surviving satellites, while the median $|v_{\rm rad,in}/v_{\rm tot,in}|$ is indeed higher for quenched satellites compared to star-forming satellites in our simulations, mass at infall is still the primary indicator for star formation or quenching.

### 3.4. Mass Loss

Disrupted satellites in our simulations underwent mass loss until they were no longer identified by the halo finder. Surviving satellites are also prone to mass loss during the time they spend in the host halo. We further quantify this in Figure 7, comparing the total virial, stellar, and gas mass lost since infall against the amount of time a satellite has spent in the host halo. Point sizes scale with infall stellar mass, and include all Mint (top row) and Near Mint (bottom row) satellites.

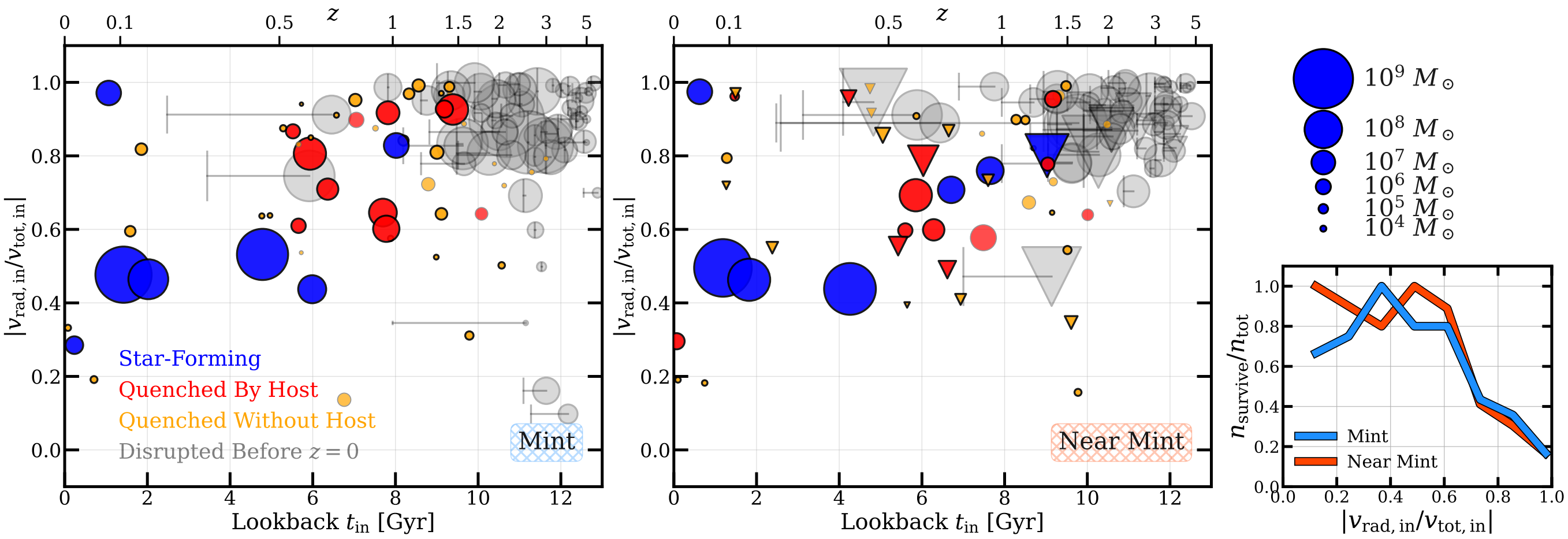

**Figure 6.** Left: the fraction of total velocity in the radial direction at infall, $v_{rad,in}/v_{tot,in}$, for satellites as a function of their lookback time to infall, for Mint satellites. Point color and styles are as in Figure 3. The size of each point scales logarithmically with the stellar mass of the satellite at infall as shown to the right, with more massive satellites in larger markers. Horizontal lines out to the time of disruption are included for disrupted satellites. Center: similar to the left panel, for Near Mint satellites, where circles are used for satellites of `Sandra` and `Elena`, and triangles for `Ruth` and `Sonia`. Right: the fraction of total Mint (light blue) and Near Mint (orange) satellites that survive, binned as a function of $v_{rad,in}/v_{tot,in}$.

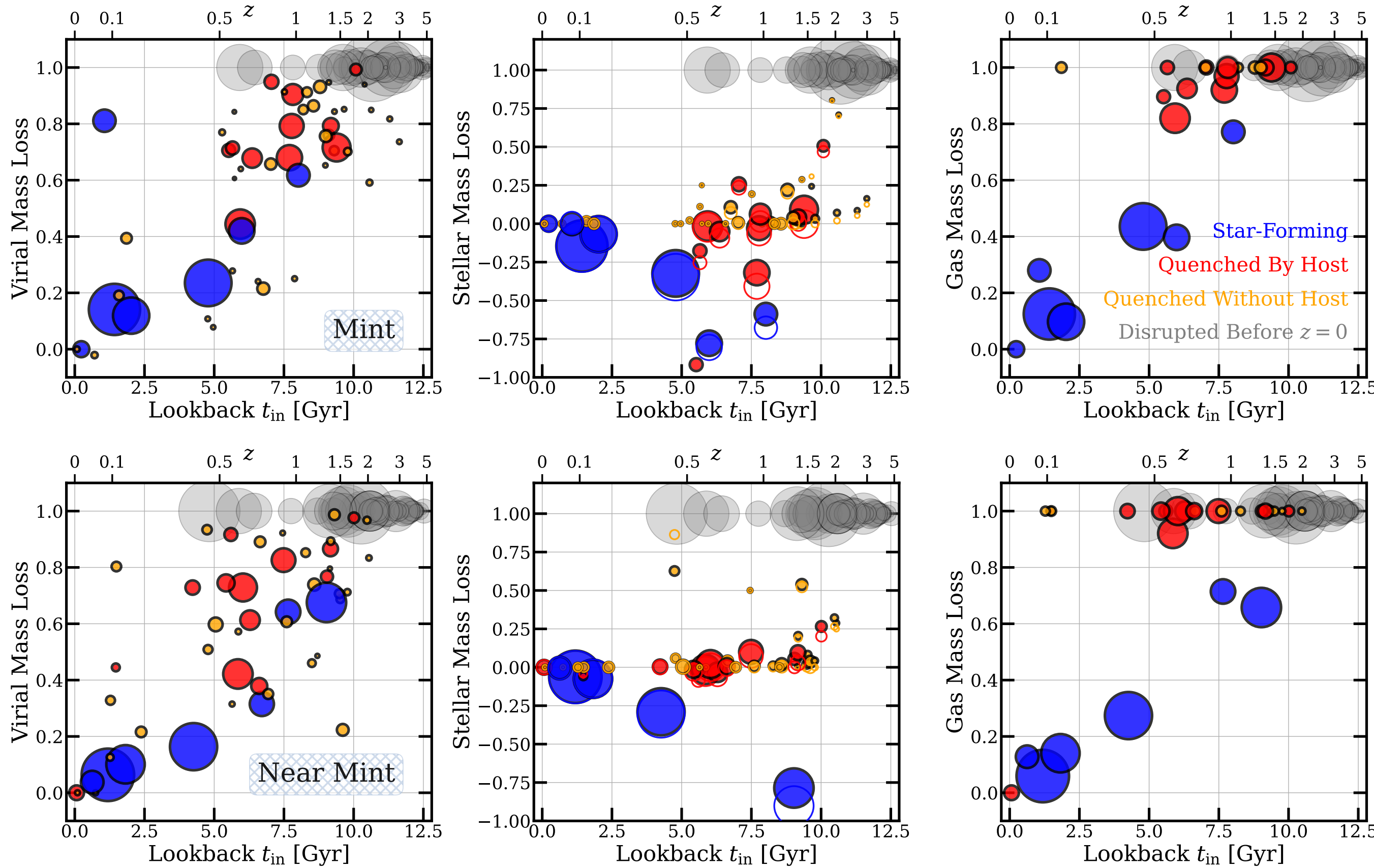

**Figure 7.** Top left: fraction of virial mass lost since infall as a function of lookback time to infall for Mint satellites. Marker colors and sizes are as in Figure 6, where disrupted satellites have mass-loss fractions of one. Top middle: fraction of stellar mass lost since infall as a function of lookback time to infall for Mint satellites. Fractions of star particles lost since infall (which includes mass loss due to stellar evolution) are indicated in the corresponding open circles. Negative values indicate a growth in stellar mass since infall. Top right: fraction of gas mass lost since infall as a function of lookback time to infall for Mint satellites. Many satellites infell with no gas, and are hence excluded from this panel. Bottom row: similar to the panels in the top row, but for all Near Mint satellites for comparison.

The amount of virial mass lost since infall scales with the time since infall, with some scatter. Massive star-forming satellites are most resistant to virial mass loss since they fell in more recently and are on more circular orbits, while less massive quenched satellites are more prone to losing dark matter from their halo with each pericentric passage. Much of the scatter in the fraction of $M_{\rm vir}$ lost among quenched satellites is explained by the variation in the number of pericentric passages completed by each satellite (see, e.g., Figure 5). With each pericentric passage, these satellites become less dark-matter dominated as they lose virial mass more rapidly than stellar mass.

Stellar mass loss proceeds more slowly for both low- and high-mass satellites. Most surviving massive satellites continue to actively grow their stellar mass after infall, typically forming up to 30%–40% of their $z = 0$ stellar mass while within the host halo. A few intermediate-mass satellites also show signatures of triggered star formation followed by rapid quenching. The Mint resolution simulations also include an outlier satellite that assembles close to 90% of its $z = 0$ stellar mass in an ∼500 Myr burst of star formation triggered at infall. On the other hand, low-mass satellites are more prone to losing stellar mass during close encounters with the host, albeit at much slower rates than their virial mass loss. Figure 7 includes both the change in total stellar mass (filled points) and number of star particles at infall (empty circles) to account for mass loss due to stellar evolution. The loss in star particles from low- and intermediate-mass satellites may be observed as stellar streams during the disruption process (see, e.g., N. Panithanpaisal et al. 2021; N. Shipp et al. 2023).

Finally, ram pressure stripping allows for gas to be removed from satellites on much shorter timescales than stellar or virial mass is stripped. Most quenched satellites fell in with no gas particles and remain gas poor, and are absent from the right panels of Figure 7. Intermediate-mass satellites that infall with gas reservoirs lost their gas rapidly after infall. Hence, the vast majority of $z = 0$ surviving satellites are currently in the process of losing dark matter and gas, while star particles are lost more slowly, and the most massive surviving satellites continue forming new stars.

## 4. Discussion

### 4.1. Role of Resolution

Simulated satellites can be prone to artificial disruption in lower-resolution realizations of baryonic mass simulations due to insufficient baryonic mass resolution (F. Jiang et al. 2021, and references therein). Recent high-resolution simulations (e.g., R. J. J. Grand et al. 2021) and semianalytic models of satellite galaxy evolution have revealed that artificial satellite disruption is mostly numerical in origin, and is caused primarily by inadequate force softening and runaway instabilities introduced in the simulations by the amplification of discreteness noise in the presence of a tidal field (e.g., F. C. van den Bosch et al. 2018; F. C. van den Bosch & G. Ogiya 2018; F. Jiang et al. 2021). E. Applebaum et al. (2021) find overall convergence in host and satellite properties between the Near Mint (high) and Mint (highest) resolution versions of the D.C. JUSTICE LEAGUE simulations used in this paper, which reach resolutions comparable to the Level 3 and Level 2 runs presented in R. J. J. Grand et al. (2021). The inclusion of halos from both these resolutions in this paper allows us to further examine the impact of resolution on these results. A comparison between halos in the Near Mint and Mint runs of each system shows that those in the intermediate- to high-mass range ($M_* > 10^6\ M_\odot$) have similar stellar masses, halo masses, infall times, survival status, and quenched status. As evident in Figure 3, satellites from the Near Mint simulations tend to be accreted slightly (<500 Myr) later, and in one case, a satellite that is in the process of quenching is still forming stars in the Mint version. Taken broadly, however, both resolutions show the same trends in mass and timescales of infall and quenching, and have found a nearly 1:1 correlation between large-scale properties such as infall times, masses, SFRs, and quenching timescales across both realizations for satellites in the intermediate- to high-mass range ($M_* > 10^6\ M_\odot$). The orbital properties at infall are more subject to changes across resolution, as can be seen in Figure 6. Beyond the effect of resolution on dynamical friction, this variation is likely because orbital dynamics can be especially sensitive to small differences due to stochasticity across realizations. A stellar mass of $10^4\ M_\odot$ is the lower limit for resolved satellites in the Near Mint simulations, and any crossmatching becomes difficult to perform for galaxies around this stellar mass because of stochasticity in the infall times and orbital trajectories after infall for such low-mass systems.

Future simulations with higher mass resolution combined with improved feedback recipes that can reproduce the chemical abundance patterns in UFD galaxies will be crucial for complementing discoveries from upcoming surveys with high angular resolution that can probe fainter signatures of disruption in low-mass systems.

### 4.2. Definitions of Disruption

The definition of disruption is dependent on methodology, with observers typically using very different definitions than simulators. These differences mean that additional nuance is required when comparing the results from our simulations to observations. Here, disruption is taken to occur when the halo finder is no longer able to identify the halo, and the majority of the particles previously associated with the satellite are identified as a part of the main halo in subsequent time steps. Disruption is then confirmed by checking the mass-loss history of each disrupted satellite. Since we require that each identified halo contains at least 100 particles (total number across all particle types), any halo that falls below that amount is considered disrupted, which results in an effective total mass floor of about $1.79 \times 10^6\ M_\odot$ ($4.2 \times 10^6\ M_\odot$) in the Mint (Near Mint) simulations, depending on the relative amounts of star and dark-matter particles contained. However, most satellites that disrupt in our simulations are in the intermediate- to high-mass range. For these satellites, it turns out that disruption generally proceeds so rapidly that the exact particle count threshold used does not have much impact on the disruption timescale relative to our time step spacing of ∼250 Myr. This is because disruption, when it happens for massive satellites in our simulations, generally occurs rapidly within $\lesssim 2$ Gyr, rather than through the gradual tidal stripping of material. When testing our analysis by increasing the minimum particle number from the fiducial particle count of 100 gradually to 1000, we found that this made no impact on the timescales of disruption for intermediate- and high-mass satellites.

Where we do see differences is in the identification and survival of low-mass halos. Requiring a higher particle threshold for halo identification reduces the number of UFDs identified, and increasing the threshold beyond $n = 1000$ effectively raises the mass threshold for satellite identification. Additionally, when we increase the minimum particle count gradually from the fiducial $n = 100$ to $n = 250$, 500, and 1000, the resulting survival fraction for UFDs decreases from 94% at $n = 100$ to 86% at $n = 250$, 74% at $n = 500$, and 60% at $n = 1000$. For those 3–11 additional galaxies that are classified as "disrupted" when increasing the minimum particle threshold, each factor of 2 higher threshold shortens the disruption times for UFD galaxies by 1–2 Gyr.

An additional factor influencing the disruption timescales is the temporal spacing of the snapshots. These simulations have typical snapshot spacings of 300 Myr. Although star formation is recorded between snapshots and orbital parameters are interpolated across snapshots, the identification of a halo as "disrupted" only occurs at discrete snapshots. As a result, the disruption times are only known to within ∼300 Myr.

Additionally, the spacing of the snapshots may influence the robustness of the halo tracing, especially during rapid pericentric passage of satellites through high-density regions close to the host halo. This was the case for ∼10% of our most massive satellites $M_{*,\rm in} \geqslant 10^8$ $M_\odot$, where AHF stopped identifying the substructure as a satellite for one time step during a pericentric passage through the dense disk of the host, but successfully reidentified the halo as it emerged from the disk. The formation histories of this subset of satellites were patched separately after confirming their orbital kinematics and mass-loss histories and included in the sample. In addition, while the orbital trajectories of UFDs are not uniformly radial, as shown in Figures 5 and 6, many UFDs (including the two that are disrupted) and most disrupted satellites show primarily radial orbits. From Figure 6, AHF can successfully identify and track UFDs on highly radial orbits at infall in our sample throughout several pericentric passages. While it is possible that some low-mass halos on rapid radial orbits could be lost by AHF in high-density regions, we checked for (using tracks such as in Figure 5) and corrected instances of high-mass halos being temporarily "lost," which we patched by hand, but were unable to find low-mass "lost" halos in our sample. However, it is possible that a different halo finder or a smaller snapshot spacing may result in slightly different progenitor tracks.

Halo finder algorithms can have difficulties identifying disrupting satellite galaxies, in particular, they struggle with tidal tails and other signatures of disruption that show nonspherical geometries (B. Diemer et al. 2024). As a result, the halo finder may report a kinematically coherent structure as "disrupted," resulting in shorter disruption timescales than other definitions. More computationally expensive algorithms that follow subhalos by tracking particle kinematics across time steps may perform better on tidally distorted halos (see, e.g., B. Diemer 2018; B. Diemer et al. 2024). The impact of adopting such an algorithm could be longer disruption times and fewer disrupted galaxies, although we note that it would likely have little effect on the low-mass end, given the small number of low-mass disrupted satellites identified in this work.

In addition to tidal streams and other kinematically coherent structures being classified as disrupted, it is also possible that "surviving" satellites may show evidence of tidal disruption. As a result, some of the satellites considered "surviving" here might be most similar to the halos identified as halo substructure by R. P. Naidu et al. (2022) and other observational work. For this reason, simulation work by N. Panithanpaisal et al. (2021), N. Shipp et al. (2023), and D. Horta et al. (2023) separates progenitors into "satellites," "tidal tails," and "phase mixed," rather than just surviving or disrupted. While we leave such a comparison for future work, we quantify the fractional change in the total and stellar mass due to mass loss of the satellites since their first infall (Figure 7) and indicate throughout the paper galaxies that have lost at least 5% of their star particles since infall. As discussed in Section 3.4, even among the surviving galaxies, most have lost large fractions of their total mass since infall. Satellites are very sensitive to virial mass loss, including a star-forming galaxy that infell 1 Gyr ago but lost ∼80% of its dark matter due to a highly radial orbital trajectory at infall. Even more indicative is the change in the stellar mass. As discussed in Sections 3.3 and 3.4, many of the star-forming galaxies have substantial increases in their stellar mass since infall, even while dark-matter mass is being lost. Similar behavior was observed in N. Panithanpaisal et al. (2021), where the time of peak stellar mass occurred a couple of billion years after the accretion time for galaxies with $M_* \gtrsim 10^8$ $M_\odot$ and some continued to form stars, even as a stellar stream formed. In contrast, most (but not all) galaxies quenched at $z = 0$ end with lower stellar masses than they started with. Some of this stellar mass loss is from stellar evolution, but some is due to the actual stripping of star particles, as can be seen by comparing the fractional reduction in stellar mass to the fractional reduction in the number of star particles from Figure 7. Different definitions can effect survival rates; for example, the survival rate for UFDs accreted up to 12 Gyr ago drops from 94% to 60% when galaxies that lose >5% of their stars are considered disrupted. Future work will provide a detailed accounting of the balance between new star particles formed and those being stripped during disruption.

### *4.3. Comparison with Other Measurements of Disruption and Quenching Timescales*

Consistent with our analysis, whether and on what timescale an accreted subhalo is disrupted depends on its time of accretion, mass, and orbital properties. More massive satellites have shorter disruption timescales due to faster orbital decay from increased dynamical friction and faster tidal disruption due to the strong influence of the host's gravitational potential (e.g., J. S. Bullock & K. V. Johnston 2005; K. V. Johnston et al. 2008; F. C. van den Bosch et al. 2016; N. C. Amorisco 2017; R. D'Souza & E. F. Bell 2021). The baryonic mass of a satellite at infall (e.g., S. Geen et al. 2013) and its time of infall (e.g., A. Fattahi et al. 2020) have also been identified as strong predictors for whether a satellite survives at $z = 0$ (J. S. Bullock & K. V. Johnston 2005; K. V. Johnston et al. 2008; F. C. van den Bosch et al. 2016; N. C. Amorisco 2017; Y. M. Bahé et al. 2019). Combined with the effect of dynamical friction, other simulations have shown that more massive satellites disrupt faster through tidal stripping of both baryonic and dark matter (Y. M. Bahé et al. 2019; E. N. Kirby et al. 2020; A. Merritt et al. 2020; A. D. Montero-Dorta et al. 2024). We also see significantly higher rates of disruption for the most massive satellites. Only 20% of all accreted satellites with $M_{*,\rm in} \geqslant 10^8$ $M_\odot$ survive at $z = 0$. Unlike A. Fattahi et al. (2020), though, who also observed increased rates of disruption for galaxies with $M_* \lesssim 10^6$ $M_\odot$, we find that the likelihood of disruption continues to decrease along with stellar mass. In the Mint resolution simulations, 84% of satellites with $M_{*,\rm in} < 10^6$ $M_\odot$ survive.

Close encounters with the host's baryonic disk also speed up satellite disruption (K. V. Johnston et al. 1995; A. M. Brooks & A. Zolotov 2014), while the presence of a supernova-generated cored profile or a disk makes a satellite more susceptible to tidal disruption (see, e.g., J. Peñarrubia et al. 2010; A. M. Brooks & A. Zolotov 2014; R. Errani et al. 2017, 2023). Notably, extremely few galaxies in this simulation with $M_* \geqslant 10^6\, M_\odot$ are identified at $z = 0$ by the halo finder as a coherent object while also having lost a significant fraction (>5%) of their infall stellar mass—a combination of characteristics that could indicate stream formation. In contrast, N. Panithanpaisal et al. (2021) found comparable numbers of stream progenitors as satellites across the entire mass range of $6 \leqslant \log(M_*/M_\odot) \leqslant 8.5$. To fully investigate these differences between simulations will require implementing a stream-identification method for these simulations (J. Guerra et al. 2025, in preparation); however, it is plausible that differences in the stellar feedback and ISM models cause the UFDs in these simulations to be more or less resistant to disruption. Such an analysis of stream formation, therefore, may offer an additional metric for constraining feedback.

We find disruption timescales on the order of a snapshot spacing (∼300 Myr) for progenitors accreted more than 10 Gyr ago. Disruption times for progenitors accreted after that time are much longer, extending up to ∼4 Gyr. These timescales are consistent with S. E. Grimozzi et al. (2024), although among their larger sample of galaxies, they found some with disruption times up to 9–10 Gyr. In comparison, N. Panithanpaisal et al. (2021) found that stellar streams formed between 0 and 5 Gyr after infall, with the time until complete disruption taking even longer. We also find that surviving satellites have a median infall lookback time of 7 Gyr, while disrupted satellites have a median infall lookback time of 11 Gyr, as included in Table 2. These times are roughly consistent with A. Fattahi et al. (2020), who found median lookback infall times of ∼7 Gyr for surviving satellites, ranging from ∼8 Gyr for satellites with $M_* \sim 10^6\, M_\odot$ to ∼4 Gyr for satellites with $M_* \sim 10^9\, M_\odot$.

The initial conditions used for these simulations were selected to produce hosts that had no major mergers since $z = 0.5$ and to produce a range of host masses, halo spins, and local densities that were still considered consistent with the Milky Way's local environment (the two Mint simulations representing the low- and high-mass ends of this range). However, with only four host halo simulations, it is impossible to fully sample the range of assembly histories. Additionally, the assembly histories of galaxy pairs (such as the M31 and Milky Way) may be weighted toward earlier times than the isolated Milky Way–mass galaxies analyzed here (I. B. Santistevan et al. 2020). Some aspects of the Milky Way's assembly history are more common in our simulations–for example, 2/4 of the hosts experienced a major merger around $z \sim 2$. On the other hand, only one of the halos, `Sandra`, experiences the late accretion of Magellanic Cloud–mass satellites.

Additionally, all of the simulations have relatively quiet accretion histories between 2 and 4 Gyr ago (see Figure 3). These quiet periods are likely the result of random sampling combined with the overall decrease in merger rates over time, and they are consistent with the Milky Way's accretion history, to the extent it is known. The outcome of these merger histories, though, is fewer high-mass, late accreted satellites from which to draw conclusions. It is possible that a larger number of such satellites would have resulted in some with the high-[Fe/H] but low-[$\alpha$/Fe] abundances observed for the Magellanic Clouds and Fornax. However, given that the surviving satellites of similar stellar mass and [Fe/H] produced in these simulations tend to have higher values of [$\alpha$/Fe] than observed, the discrepancy may instead lie in either the SFHs or enrichment models.

The quenching timescales of satellites around Milky Way–like galaxies have also been studied through a combination of recent observational and theoretical studies. Observed low- and intermediate-mass ($M_* < 10^8\, M_\odot$) satellites have higher quenched fractions and faster quenching times (∼1–3 Gyr) than massive satellites ($M_* \geqslant 10^8\, M_\odot$; S. P. Fillingham et al. 2015; D. R. Weisz et al. 2015; A. R. Wetzel et al. 2015). Observations suggest that galaxies in high-density environments can quench rapidly (within ∼1 Gyr) after their first pericentric passage through the host halo (A. K. Upadhyay et al. 2021). Simulations by C. M. Simpson et al. (2018) have found that half of the satellite galaxies quench within 1 Gyr of infall, with the remainder within 5 Gyr. Similar results were found by J. Samuel et al. (2022) for the FIRE simulations. The quenching timescales we detect here, shown in Figure 4 and explored in greater detail for the Near Mint subsample in H. B. Akins et al. (2021), are generally within about 2 Gyr of infall, although others can take as long as 5 Gyr, which is consistent with both observations and other simulations.

### 4.4. Observational Implications

#### 4.4.1. Metallicities

Additional observational metallicity information will be crucial for checking whether the bimodal metallicity trend seen between surviving and disrupted satellites (see Figure 1) persists for galaxies with $M_* < 10^6\, M_\odot$. R. P. Naidu et al. (2022) do not observe satellites below this mass range, and none of the simulated satellites below this mass threshold disrupt despite showing a range of infall parameters. Many of these satellites complete multiple pericentric passages and remain on stable eccentric orbits without disruption at $z = 0$. Even when examining the fraction of star particles lost as a proxy for disruption, we see no evidence for this bimodality extending down to the ultrafaint regime, as all UFDs, surviving or not, are quenched by reionization.

Looking for signatures of disruption, such as tidal stripping in these UFDs will provide important context for the survival rates of low-mass satellites. Upcoming observational advancements in large-area deep surveys such as the Vera C. Rubin Observatory and Nancy Grace Roman Space Telescope (Roman) will be crucial for increasing the census of known UFDs over a range of stellar masses (e.g., B. Mutlu-Pakdil et al. 2021). Combined with follow-up observations at high angular resolution through the Hubble Space Telescope, JWST, and Roman, better constraints on the chemical abundances, SFHs, and kinematics will enable improved cross correlation of the chemical enrichment, SFH, and stellar populations of nearby UFDs with simulated UFDs.

#### 4.4.2. Variation in Surviving Satellite Populations across Hosts

There are a wide variety of possible explanations for the observed differences in the quenched fraction of Local Group satellites with $10^{7.5}\, M_\odot \lesssim M_* \lesssim 10^{8.5}\, M_\odot$ and those observed by the SAGA and ELVES surveys. These possible explanations include difficulties in correcting for the observational bias toward bluer, star-forming satellite galaxies, especially in the earlier releases (A. S. Font et al. 2022), and differences in mass and environment between the typical hosts in the SAGA and ELVES survey and the Milky Way and Andromeda (e.g., A. S. Font et al. 2022; J. Samuel et al. 2022; C. Engler et al. 2023; J. Van Nest et al. 2023). Given that the Milky Way quenched fractions lie within the spread of SAGA hosts, though, it is possible that there is simply a large amount of stochasticity in the quenched fractions and that the Milky Way and Andromeda are simply somewhat unusual (Y.-Y. Mao et al. 2024). Some possible sources of variation in quenched fractions are the concentration of the dark-matter halo of the host (e.g., C. E. Fielder et al. 2019), and differences in the accretion histories of the host galaxy. Variations in satellite infall time, along with differences in the disruption time due to satellite mass and orbit, change the number of surviving satellites, thus altering the denominator of the quenched fraction.

In the mass range $10^7\ M_\odot \lesssim M_* \lesssim 10^8\ M_\odot$, where the Local Group exhibits substantially higher quenched fractions than the median SAGA and ELVES hosts, the quenching timescales range from a couple of gigayears before infall to 5 Gyr after (H. B. Akins et al. 2021). Here, we show disruption timescales in this mass range that similarly vary from 0–5 Gyr, depending on the time and trajectory of infall. Since almost all disrupted galaxies in this mass range were accreted prior to 10 Gyr ago, the specific history of infall will affect the number of surviving quenched halos. For example, because of the role of disruption in setting the denominator of the quenched fraction, we find that a larger fraction of galaxies in this mass range accreting somewhat more recently than 10 Gyr ago could result in a larger number of quenched halos at $z = 0$ and, consequentially, a higher quenched fraction. Therefore, the Milky Way's relatively high quenched fraction of classical dwarf galaxies can be explained if many of the galaxies in this mass range were accreted between 5 and 10 Gyr ago. At the opposite end of the mass spectrum, the fact that both of the Milky Way's bright dwarf satellites ($\log(M_*/M_\odot) \gtrsim 8$) are star forming is consistent with the Milky Way accreting this mass range of galaxies either relatively recently (allowing them to survive as star-forming satellites, such as the LMC and SMC) or sufficiently long ago that they were able to disrupt (such as GSE). This halo accretion history is consistent with the interpretation of Y.-Y. Mao et al. (2024) that the Milky Way is an older, less massive host with a recently accreted LMC/SMC system.

Since satellites are often accreted together (R. D'Souza & E. F. Bell 2021), variations in merger histories can have an outsized effect on the populations of surviving satellites. Furthermore, the characteristics of low-mass satellites can be used to infer the properties of the higher-mass satellites they may have accompanied. As evident in Figure 3, low-mass satellites are more likely to survive for longer periods than higher-mass satellites, and classical dwarfs with $10^6\ M_\odot < M_* < 10^8\ M_\odot$ tend to quench within 1–2 Gyr of first accretion. As a result, they can be good tracers of more massive accretion events. R. D'Souza & E. F. Bell (2021) use this same line of reasoning to posit a relatively early accretion of M33 onto M31 (∼4–9 Gyr ago) based on the lack of recently quenched satellites of M31. An additional complication when considering group accretion, though, is that low-mass galaxies in groups may experience preprocessing. Therefore, the time between when a low-mass galaxies crosses the virial radii of the host and when it quenches is likely to be shorter or even negative (J. Samuel et al. 2022).

## 5. Summary

In this paper, we compare the timescales of quenching and disruption for the full census of accreted satellites of simulated Milky Way analogs that include both surviving and disrupted satellites. The timescales of quenching and disruption compete to produce the $z = 0$ observed star-forming and quenched satellite populations. Our main findings are as follows.

1. Our simulations successfully reproduce the observed trends from S. P. Fillingham et al. (2019) and R. P. Naidu et al. (2022) in stellar mass–[Fe/H]–[$\alpha$/Fe] between surviving and disrupted satellites (Section 3.1). Surviving satellites generally show higher [Fe/H] and lower [$\alpha$/Fe] than their disrupted counterparts in the same mass range.
2. The timescales of quenching and disruption for satellites are strongly correlated with their mass at infall and lookback time to infall (Section 3.2, Table 2).
    1. Most ultralow- and low-mass satellites ($M_{*,\rm in} \leqslant 10^6\ M_\odot$) quench $>2 \times R_{\rm vir}$ away from the host and more than 1 Gyr prior to infall, likely at reionization. These satellites infall at a range of cosmic times and have typically completed several pericentric passages before $z = 0$. Of the low-mass satellites accreted within the last 12 Gyr, 94% survived, although most lose significant portions of their virial mass.
    2. Intermediate-mass classical dwarf galaxies ($10^6\ M_\odot < M_{*,\rm in} \leqslant 10^8\ M_\odot$) are sensitive to quenching by the host halo environment. The timescale of quenching is short, and these satellites typically quench within ∼1–3 Gyr of infall. Disruption is likely for galaxies in this mass range that accreted more than 8 Gyr ago. Disrupted galaxies in this mass range remained star forming up until their time of disruption.
    3. High-mass satellites ($M_{*,\rm in} \geqslant 10^8\ M_\odot$) are more resistant to quenching by the host halo, and remain star forming for several gigayears after infall. However, satellites in this mass range are more prone to disruption, and do not survive more than ∼5 Gyr after infall.
    4. Most satellites accreted more than 11 Gyr ago are disrupted by the host halo, independent of their mass at infall. 92% of all disrupted dwarfs remained star forming up until disruption.
3. Orbital trajectories at infall have a secondary influence on quenching and disruption. Most massive satellites ($M_{*,\rm in} \geqslant 10^8\ M_\odot$) that manage to survive and remain star forming at $z = 0$ infell with and maintain more circular orbits (Section 3.3).
4. Disruption is a more rapid process than quenching for satellites with $M_{*,\rm in} \geqslant 10^6\ M_\odot$, while quenching proceeds more rapidly than disruption for low-mass satellites. All accreted satellites lose dark matter and gas faster than stars, and can lose up to ∼80% of their dark matter and gas since infall while still retaining or even growing their stellar mass through active star formation after accretion (Section 3.4).

## Acknowledgments

We thank the anonymous referee for the valuable feedback. We are also grateful to Shonda Kuiper for her assistance in the statistical analysis. D.P. acknowledges support from the NSF GRFP and from Grinnell College through the Mentored Advanced Project program. D.P. and C.C. were also supported by the NSF under CAREER grant AST-1848107. Resources supporting this work were provided by the NASA High-End Computing (HEC) Program through the NASA Advanced Supercomputing (NAS) Division at Ames Research Center. This research is part of the Blue Waters sustained-petascale computing project, which is supported by the National Science Foundation (awards OCI-0725070 and ACI-1238993) and the state of Illinois. Blue Waters is a joint effort of the University of Illinois at Urbana-Champaign and its National Center for Supercomputing Applications. This research also used the NSF-supported Frontera project operated by the Texas Advanced Computing Center (TACC) at the University of Texas at Austin.

## ORCID iDs

Debosmita Pathak https://orcid.org/0000-0003-2721-487X
Charlotte R. Christensen https://orcid.org/0000-0001-6779-3429
Alyson M. Brooks https://orcid.org/0000-0002-0372-3736
Ferah Munshi https://orcid.org/0000-0002-9581-0297
Anna C. Wright https://orcid.org/0000-0002-1685-5818
Courtney Carter https://orcid.org/0000-0003-1144-7433